\newcommand{\Ho}{Ho$_{2}$Ti$_{2}$O$_{7}$} 
\newcommand{\Dy}{Dy$_{2}$Ti$_{2}$O$_{7}$}
\newcommand{\Tb}{Tb$_{2}$Ti$_{2}$O$_{7}$}

\newcommand{\Is}{\ensuremath{\langle 111 \rangle}}
\newcommand{\e}{{\rm e}}

\documentclass[twocolumn,showpacs,preprintnumbers,amsmath,amssymb,prb]{revtex4}
\usepackage{graphicx}% Include figure files
\usepackage{dcolumn}% Align table columns on decimal point
\usepackage{bm}% bold math

\begin{document}

\preprint{}

\title{Theory of paramagnetic scattering in highly frustrated magnets 
with long-range dipole-dipole interactions: The case of the \Tb\,  
pyrochlore antiferromagnet}

\author{Matthew Enjalran$^1$}
 \email{enjalran@gandalf.uwaterloo.ca. Current address: Department of Physics,
Southern Connecticut State University, 501 Crescent Street, New Haven, CT 06515.}  
\author{Michel J.P. Gingras$^{1,2}$}
 \email{gingras@gandalf.uwaterloo.ca}
\affiliation{$^1$Department of Physics, University of Waterloo, 
Ontario, N2L 3G1, Canada \\ 
$^2$Canadian Institute for Advanced Research, 180 Dundas Street West,
Toronto, Ontario, M5G 1Z8,Canada}

\date{\today}

\begin{abstract}

Highly frustrated antiferromagnets composed of magnetic rare-earth 
 moments are currently attracting much experimental and theoretical interest.
Rare-earth ions generally have small exchange interactions and large
magnetic moments. This makes it necessary to understand in detail the role of
long-range magnetic dipole-dipole interactions in these systems, in particular
in the context of spin-spin correlations that develop in the paramagnetic phase,
but are often unable to condense into a conventional long-range magnetic ordered phase.
This scenario is most dramatically emphasized in the
 frustrated pyrochlore antiferromagnet material
\Tb\ which does not order down to 50 mK despite an antiferromagnetic
Curie-Weiss temperature $T_{\rm CW} \sim -20$ K.
In this paper we report results from mean-field theory calculations 
of the paramagnetic elastic 
neutron-scattering in highly frustrated magnetic systems with long-range dipole-dipole
interactions, focusing on the \Tb\ system.
Modeling \Tb\ as an antiferromagnetic \Is\ Ising pyrochlore, we find
that the mean-field paramagnetic scattering is inconsistent with the
experimentally observed results. Through simple symmetry arguments
we demonstrate that the observed paramagnetic correlations in \Tb\ are 
precluded from being generated by any spin Hamiltonian that considers
only Ising spins, but are qualitatively consistent with Heisenberg-like moments.
Explicit calculations of the paramagnetic
scattering pattern for both  \Is\ Ising and Heisenberg models, which 
include finite single-ion anisotropy, support these claims. 
We offer suggestions for reconciling the need to restore spin isotropy 
with the Ising like structure suggested by the single-ion properties of Tb$^{3+}$. 

\end{abstract}

\maketitle

%%%%%%%%%%%%%%%%%%%%%%%%%%%%%%%%%%%%%
\section{Introduction \protect}
\label{sec:intro}

\begin{figure}[!ht]
\begin{center}
\includegraphics[width=3.0in,height=2.8in]{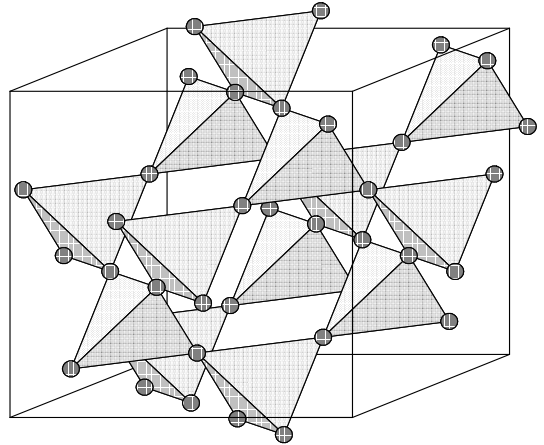}
\caption{The pyrochlore lattice structure. The lattice
is constructed from a network of corner sharing tetrahedra in which
each tetrahedron resides at an fcc Bravais lattice point. A cubic 
cell contains 4 tetrahedra or 16 atoms.}
\label{fig-pyro}
\end{center}
\end{figure}

The pyrochlore oxides, with the general formula (A$_2$B$_2$O$_7$), have attracted a 
great deal of attention over the last decade because the combination of lattice
geometry and chemical composition allow for a plethora of interesting 
physical phenomena in these materials \cite{rev-ramirez,rev-greedan,rev-SI}. 
The A and B sites reside on two distinct interpenetrating 
pyrochlore networks of corner sharing tetrahedra,
Fig.~\ref{fig-pyro}. Since A$^{3+}$ can be either a magnetic 
rare-earth or a non-magnetic transition metal, and
B$^{4+}$ can be a transition metal with or without a moment, there are numerous 
possibilities to study insulating and itinerant magnetic models in a geometrically 
frustrated environment. Experimentally, long-range magnetic ordered states 
\cite{gdtio-raju,gdtio-palmerchalker,gdtio-bramwell}, 
novel magnetic phases (e.g., 
spin-glass \cite{ymoo-tbmoo,ymoo-gingras,ymoo-gardner,tbmoo-gaulin,ymno-reimers}, 
spin-ice \cite{rev-SI,1stSI,ramirez-nature,shastry-private},  
spin liquid \cite{tbtio-gardner1,gardner-tbtio50}), 
anomalous Hall effect \cite{Anom-hall}, 
metallic properties \cite{maeno-iridates}, 
and superconductivity\cite{pyroSC1a,pyroSC1b} have been observed in the
pyrochlore oxides. Materials with the pyrochlore-related spinel structure have 
also attracted much attention recently. Heavy fermion physics has been 
observed in the $d$-electron LiV$_2$O$_4$ compound\cite{liv2o4-HF}. 
In the spinel antiferromagnet ZnCr$_2$F$_4$, a spin-Peierls-like 
transition has been observed \cite{shlee-prl} at low temperatures 
and a protectorate of weakly interacting spin directors slightly 
above the spin-Peierls transition temperature \cite{shlee-nature}.    

In insulating rare-earth magnetic pyrochlores, the
magnetic rare-earth ions often have large dipole moments, i.e., $\mu >> 1\mu_{\rm B}$
and small Heisenberg exchange interactions. In such a situation, 
long-range dipole-dipole interactions are a significant contribution to the Hamiltonian. 
As well, a large single-ion anisotropy interaction is also often present.
For example, crystal fields produce an effective Ising doublet at the rare-earth ion site,
A$^{3+}$, in the  
Ho$_2$Ti$_2$O$_7$ (Ref.~\onlinecite{hotio-rosenkranz}),
Dy$_2$Ti$_2$O$_7$ (Ref.~\onlinecite{hotio-rosenkranz}),
Tb$_2$Ti$_2$O$_7$ (Refs.~\onlinecite{hotio-rosenkranz} and \onlinecite{tbtio-gingras1}) and
Yb$_2$Ti$_2$O$_7$ (Ref.~\onlinecite{ybtio-hodges}) materials.
An energy gap separates the ground state doublet from the lowest lying excited states 
with the Ising quantization axis coinciding with the local cubic \Is\ directions 
\cite{hotio-rosenkranz,tbtio-gingras1}, i.e., the quantization 
axis points toward the center of the tetrahedral basis unit cell. 
Most noticeably, this strong local \Is\ Ising axis anisotropy 
has been found responsible for endowing frustration to the pyrochlore 
lattice in the presence of effective nearest-neighbor 
ferromagnetic (FM) interactions \cite{1stSI,bramwell-jpc,111-moessner-prb}. 
From a statistical mechanics point of view, a FM \Is\ Ising pyrochlore model is 
equivalent to a model for disordered water ice, $I_h$, where both 
magnetic \cite{1stSI} and water ice \cite{bernal-fowler,pauling,pauling-chembond}
models possess macroscopic degeneracy. Recent experiments on the 
magnetic systems \Dy\ (Ref.~\onlinecite{ramirez-nature}) and 
\Ho\ (Refs.~\onlinecite{hotio-spincorrel} and \onlinecite{cornelius-hotio}), 
so called spin-ice materials, 
reveal a residual entropy in agreement with
the prediction for water-ice \cite{pauling-chembond}. 
In contrast,  nearest-neighbor antiferromagnetic (AFM) interactions
in a \Is\ Ising pyrochlore model are non-frustrated 
\cite{bramwell-jap,bramwell-jpc,111-moessner-prb,tbtio-gingras1};
therefore, such a model is expected to order
at a temperature set by the nearest-neighbor energy scale, 
i.e., $T_N \approx J^{\rm eff}_{\rm nn}/k_{\rm B}$. 
A candidate AFM \Is\ pyrochlore material 
is \Tb , where a non-collinear long-range ordered state 
is expected at $T_N \approx 1\, {\rm K}$ 
(Refs.~\onlinecite{tbtio-gingras1} and \onlinecite{dipSImodel1}). 
However, experiments indicate the material fails to order down to 
$T=50\, {\rm mK}$ 
(Refs.~\onlinecite{tbtio-gardner1} and \onlinecite{gardner-tbtio50}),
despite an antiferromagnetic Curie-Weis temperature, $\theta_{\rm CW}\sim -20$ K,
making this system so far one of the cleanest  realizations of a spin liquid
in a three dimensional system. 
The mechanism responsible for \Tb\ failing to order down to such low temperature
has not yet been resolved \cite{enjalran-JPC}.

In \Tb , the energy gap between the ground state doublet and the
first excited state doublet is $\Delta \approx 20$ K 
(Refs.~\onlinecite{tbtio-gingras1} and \onlinecite{hotio-rosenkranz}).
The size of anisotropy gap here is an order of magnitude
smaller than what is observed in the spin-ice materials,
$\Delta = 250-350$ K (Ref.~\onlinecite{hotio-rosenkranz}),
and is comparable to $\theta_{\rm CW}$ for \Tb . 
Consequently, in early theoretical work on \Tb\, a \Is\ Ising model was assumed and  
a non-collinear long-range ordered state with zero net moment 
about each unit tetrahedron (${\bm q} = {\bm 0}$ state) 
was predicted at $T\approx 1\; {\rm K}$
(Refs.~\onlinecite{dipSImodel1} and \onlinecite{tbtio-gingras1}).
However, muon spin relaxation measurements indicate a dynamically fluctuating 
state at all experimentally achievable temperatures, which recent results
pushed down below $50\; {\rm mK}$ (Ref.~\onlinecite{gardner-tbtio50}). 
In contrast, some static susceptibility data 
suggest a spin-glass state at these temperatures 
\cite{corruccini-tbtio}.
Despite the experimental evidence that
the magnetic single-ion ground state 
of Tb$^{3+}$ is an Ising doublet~\cite{hotio-rosenkranz,tbtio-gingras1},
several experimental groups have used 
simple Heisenberg type models (which for the pyrochlore 
lattice is paramagnetic down to $T \rightarrow 0^+ \; $,
Refs.~\onlinecite{villain,rbs-mft,moessner-chalker-prl})
to obtain qualitative agreement with the observed 
paramagnetic (PM) scattering pattern\cite{tbtio-gardner2,tbtio-yasui1}.
Finally, 
we note that high pressure neutron-scattering experiments on \Tb\ 
find a transition to long-range order at a temperature
in excess of 1 K for an applied pressure larger than 2 GPa \cite{mirebeau-Nature}, but 
the magnetic structure has not been determined.
In summary, there is currently ambiguity 
even as to the nature of the 
paramagnetic correlations developing in
\Tb\ at temperatures above $\theta_{\rm CW}$. 
This is an important issue since very recent neutron-scattering 
measurements on a single crystal of \Tb\
down to 50 mK have found that the scattered intensity in reciprocal space remains
essentially unchanged (i.e. frozen out) when going from the paramagnetic
temperature of 10 K down to 50 mK (Ref.~\onlinecite{gardner-tbtio50}).
It would therefore appear that a first 
requirement to make theoretical
progress in understanding the spin liquid state in \Tb\ would be  to understand
the nature of the paramagnetic spin$-$spin correlations.
It is the aim of this paper to shed some light on the paramagnetic 
correlations of this material.

In this article, we use mean-field theory (MFT) to investigate the paramagnetic 
spin-spin correlations of a \Is\ Ising dipolar model pertinent to \Tb\ as well
as a finite Ising anisotropy model.
In MFT, the PM regime is realized for temperatures above an
energy scale set by the mean-field (MF) critical mode, i.e., 
$T > T_c^{\rm MF} \equiv \lambda_{c}^{\rm MF}/n$,
where $T$ is temperature in units of $k_{\rm B}$,
$n$ is the number of spin components, and $\lambda_{c}^{\rm MF}$ 
represents the mean-field global maximum eigenvalue of the 
${\bm q}$-dependent susceptibility. As the temperature
approaches $T_c^{\rm MF}$, one expects the critical mode to 
control the spin-spin correlations. 
A clear understanding of this critical mode softening relies 
on an accurate treatment of all interactions in the
Hamiltonian. The long-range dipole-dipole interactions are 
our main concern. Our major results are the following:
We establish, on symmetry grounds and through calculations,
that the observed PM neutron-scattering is inconsistent with a 
local \Is\ Ising dipolar model. Calculations performed for 
an anisotropic Heisenberg pyrochlore model yield good agreement 
with experiment and thus support the claim that at least a partial 
restoration of spin isotropy occurs in \Tb .

%%%%%%%%%%%%%%%%%%%%%%%%%%
We also have in mind a broader perspective in presenting the enclosed work and
detailed derivation of the mean-field formulation of the structure factor, $S({\bf q})$, 
for highly frustrated magnets. In the past few years, 
there have been a number of interesting
and puzzling thermodynamic data and neutron-scattering 
results on highly frustrated magnets.
Examples include  the antiferromagnetic
Gd$_3$Ga$_5$O$_{12}$ garnet (referred to as GGG),
(Refs.~\onlinecite{ggg-wolf,ggg-shiffer,ggg-shiffer2,ggg-pentrenko,
ggg-pentrenko-prl,bonville-ggg})
the pyrochlore antiferromagnet Gd$_2$Ti$_2$O$_7$,
(Refs.~\onlinecite{gdtio-raju,gdtio-palmerchalker,gdtio-ramirez,
gdtio-bramwell,bramwell-gto-JPC}) 
and the pyrochlore ferromagnet Yb$_2$Ti$_2$O$_7$ 
(Refs.~\onlinecite{ybtio-hodges} and \onlinecite{yasui-yto}).
The experimental results on these systems, which will be 
discussed in more detail in Section \ref{sec:disc}, raise a common issue: 
To further our current understanding of a number of highly frustrated magnets, 
including quantum fluctuations \cite{delMaestro},
we need to have a 
clear and quantitative understanding of the predominant correlations that initially
develop out of the paramagnetic state as the materials are cooled. The mean-field
theory described herein is a first approach to formulate such a program. For concreteness,
this paper focuses on the specific cases of insulating pyrochlore oxides with local
\Is\ Ising axis anisotropy. It is straightforward to use the 
formalism herein to tackle the 
frustrated dipolar systems described above as well as others.
%%%%%%%%%%%%%%%%%%%%%%%%

The outline of the paper is as follows: In Section \ref{sec:mft} a 
condensed description of the MFT formalism of neutron-scattering for 
the anisotropic Heisenberg and \Is\ Ising pyrochlores is   
presented. We give the MFT results for the paramagnetic 
scattering of \Tb\ in Section
\ref{sec:neutron}. Our mean-field data are also compared to 
Monte Carlo results for the paramagnetic $S({\bm q})$.
Section \ref{sec:disc} 
discusses the need
to relax the \Is\ Ising constraint to allow for transverse fluctuations
in Tb$_2$Ti$_2$O$_7$.
For completeness, we include several detailed appendices. In 
Appendix \ref{append:nscatt}, the variational MFT is discussed and 
the equations for elastic neutron-scattering are derived in detail.
An alternate approach, a high-temperature
series expansion, to the equations for neutron-scattering is presented 
in Appendix \ref{append:htse}. The 
derivation of the Ewald equations for the ${\bm q}$-dependent dipole-dipole 
Hamiltonian is given in Appendix \ref{append:ewald}. 
Appendix \ref{append:symm} contains a detailed discussion of the symmetry 
allowed scattering patterns 
for \Is\ Ising and Heisenberg pyrochlores. 
We note that other authors have discussed various aspects of the mean-field theory
\cite{chaikin-lubensky,abharris-mft,reimers-mft,rbs-mft,kanda-mft-Ho} 
and Ewald \cite{born-huang,CK,AF,litbf-holmes} derivations 
presented in Appendices \ref{append:nscatt}
and \ref{append:ewald}, respectively. Our purpose here is to provide a 
self-contained reference for the application of the Ewald method within a 
mean-field formalism,
connecting at times with results from earlier work \cite{reimers-mft},
and which
will be useful to other researchers who wish to study frustrated 
magnetic systems with  
non-Bravais lattice geometries 
and allowing for an array of possible 
spin symmetries and interactions.

%%%%%%%%%%%%%%%%%%%%%%%%%%%%%%%%%%%%%%%%%%%%
\section{Mean-Field Theory of Neutron Scattering in the Pyrochlores \protect}
\label{sec:mft}

In this section, we present the models and provide an outline
of the derivation of the mean-field equations for the neutron-scattering cross section, 
$d\sigma (Q)/d\Omega$, for classical spins on the pyrochlore lattice.  
The resulting equations are only applicable in the disordered paramagnetic
regime of the model Hamiltonians. 
Our derivation is performed for the general anisotropic Heisenberg model 
because it has broad appeal to the study of many highly frustrated magnetic 
systems.
Our method is best described 
as variational mean-field theory \cite{chaikin-lubensky} (VMFT) 
(which is equivalent to a Gaussian approximation of the free energy) 
and has been used to study frustrated magnetic systems 
\cite{reimers-mft,rbs-mft,kanda-mft-Ho}. The details of VMFT and its application to
the scattering cross section are presented in Appendix~\ref{append:nscatt}. 
We remind the reader that MFT corresponds to a 
partial resummation of an infinite number of terms in 
the high-temperature series expansion
for the ${\bm q}-$dependent susceptibility, $\chi({\bf q})$. 
In particular, the 
correlations $\langle {\bm S}(0) \cdot {\bm S}(r) \rangle$ are correctly 
treated to order $\beta = 1/T$ (or $\chi({\bm q})$ to $1/T^2$) in MFT. 
This is demonstrated in Appendix~\ref{append:htse}, 
where scattering cross section equations are derived via a high 
temperature series expansion. 

%%%%%%%%%%%%%%%%%%%%%%%%%%%%
\subsection{Models}
\label{sec:mft-1}

The pyrochlore lattice, Fig.~\ref{fig-pyro}, is a non-Bravais lattice 
that we describe as a fcc lattice with a four atom unit cell. The positions
of the fcc Bravais lattice points, which coincide with a corner point
on the tetrahedral basis, are denoted by ${\bm R}_i$.
The four atoms that form the tetrahedron at each fcc point 
(and represent different sublattices) are labeled by ${\bm r}^a$. Hence, the 
position of a site in the pyrochlore lattice is given by 
${\bm R}_i^a={\bm R}_i+{\bm r}^a$. 
Table~\ref{tab-raza} lists our convention for the tetrahedral basis 
coordinates, ${\bm r}^a$. 
The most general Hamiltonian for rare-earth spins on the pyrochlore lattice 
is a Heisenberg model with nearest-neighbor exchange, dipole-dipole, and 
single-ion anisotropy (with a local \Is\ orientation) energies,
\begin{eqnarray}
\label{eq-Hheis}
& & H_{H} = - J \sum_{\langle (i,a),(j,b) \rangle}
{\bm S}_{i}^{a} \cdot {\bm S}_{j}^{b} 
- \Delta \sum_{i,a} (\hat{z}^{a} \cdot {\bm S}_{i}^{a})^{2} \\
& &+ D_{\rm dd} \sum_{(i,a)>(j,b)}
\left ( \frac{{\bm S}_{i}^{a} \cdot {\bm S}_{j}^{b}}{|{\bm R}_{ij}^{ab}|^3}
- \frac{3({\bm S}_{i}^{a} \cdot {\bm R}_{ij}^{ab})
({\bm S}_{j}^{b} \cdot {\bm R}_{ij}^{ab})}{|{\bm R}_{ij}^{ab}|^5} \right ). \nonumber
\end{eqnarray}
The unit vector $\hat{z}^a$ represents the local 
\Is\ quantization axis that points toward 
the center of a tetrahedron. Table~\ref{tab-raza} defines our convention for
$\hat{z}^a$. The spins ${\bm S}_i^a$ have unit length and full $O(3)$ symmetry, 
${\bm R}_{ij}^{ab} = {\bm R}_i^a - {\bm R}_j^b$ is the 
vector separation between spins ${\bm S}_i^a$ 
and ${\bm S}_j^b$, and $J$ and $\Delta$ define the exchange and single-ion 
energy scales, respectively. The convention established in Eq.~\ref{eq-Hheis} 
defines $J>0$ as FM and $J<0$ as AFM exchange energies. 
The dipolar energy scale is set by 
$D_{\rm dd}\equiv DR_{\rm nn}^3$, where 
\[
D=\frac{\mu_{\rm o}}{4\pi} \frac{\mu^2}{R_{\rm nn}^3} \; ,
\]
$\mu$ is the moment on the rare-earth ion, and $R_{\rm nn}$ is the nearest
neighbor distance. 

The \Is\ Ising dipolar model for the pyrochlore lattice \cite{dipSImodel1} 
is obtained by considering the limit of large Ising anisotropy  
in Eq.~\ref{eq-Hheis} ($\Delta/|J| \gg 1$ and $\Delta/D \gg 1$). 
The low energy physics of this system is modeled by the Hamiltonian, 
\begin{eqnarray}
\label{eq-Hising}  
& & H_{I} = - J \sum_{\langle (i,a),(j,b) \rangle}
(\hat{z}^{a} \cdot \hat{z}^{b}) \sigma_{i}^{a} \sigma_{j}^{b} \\
& &\! \! \! + D_{\rm dd} \! \sum_{(i,a)>(j,b)} \!
\left (\frac{(\hat{z}^{a} \cdot \hat{z}^{b})}
{|{\bm R}_{ij}^{ab}|^3}
- \frac{3(\hat{z}^{a} \cdot {\bm R}_{ij}^{ab})
(\hat{z}^{b} \cdot {\bm R}_{ij}^{ab})}
{|{\bm R}_{ij}^{ab}|^5} \right ) \sigma_{i}^{a} \sigma_{j}^{b} \nonumber
\end{eqnarray}
By low energy physics we mean that the single-ion term in Eq.~\ref{eq-Hheis} is removed 
and the spins are restricted to lie along the local \Is\ quantization
axis, i.e., ${\bm S}_{i}^{a}=\hat{z}^{a}\sigma_{i}^{a}$ with $\sigma_{i}^{a} \pm 1$.
If one were to truncate the dipolar sum in Eq.~\ref{eq-Hising} at nearest 
neighbor distances, then the following effective nearest-neighbor energy scale 
could be defined, 
\begin{equation} 
\label{eq-Jeff}
J_{\rm nn}^{\rm eff} \equiv J_{\rm nn} + D_{\rm nn} \ , 
\end{equation}
where $J_{\rm nn} = J/3$ and $D_{\rm nn} = 5D/3$.
For FM effective nearest-neighbor exchange, $J_{\rm nn}^{\rm eff} > 0$ 
setting all dipolar interactions beyond nearest-neighbor to zero,
one has the nearest-neighbor spin-ice model of Harris \textit{et al.}, 
Refs.~\onlinecite{1stSI} and \onlinecite{bramwell-jpc}. 
If the nearest-neighbor interactions $J_{\rm nn}^{\rm eff}$ are AFM, 
then the model possesses a unique ordered state
(${\bm q}={\bm 0}$, all-in all-out state) 
at temperatures on the order of 
$|J_{\rm nn}^{\rm eff}|$ 
(Refs.~\onlinecite{tbtio-gingras1,bramwell-jpc,111-moessner-prb,bramwell-jap}). 
The transition between the spin-ice and
${\bm q}={\bm 0}$ phases occurs at $J_{\rm nn}/D_{\rm nn} = -1.0$.  
When dipole-dipole sum is extended to long-range distances, 
the transition between the ${\bm q}={\bm 0}$ and the spin-ice states 
shifts to $J_{\rm nn}/D_{\rm nn} \cong -0.908$. 
Hence, long-range dipolar interactions
favor the ${\bm q}={\bm 0}$ AFM phase slightly \cite{dipSImodel1}.

\begin{table}
\caption[]{Our convention for vectors: The ${\bm r}^a$ define the basis vectors
and $\hat{z}^a$ define the local \Is\ anisotropy axes for spins
on the pyrochlore lattice. The size of the corresponding cubic cell
is given by ${\bar a}$ and contains $16$ atoms. Vectors $\hat{n}^u$ represent the 
global Cartesian basis vectors.}
\label{tab-raza}
%\vspace{0.5cm}
\begin{ruledtabular}
\begin{tabular}{cccccc}
${\bm r}^{(1)}$   & $\frac{\bar a}{4}$(0,0,0) & $\hat{z}^{(1)}$ & $\frac{1}{\sqrt{3}}$ (1,1,1) & 
$\hat{n}^{(1)}$ & (1,0,0) \\
${\bm r}^{(2)}$   & $\frac{\bar a}{4}$(1,1,0) & $\hat{z}^{(2)}$ & $\frac{1}{\sqrt{3}}$ (-1,-1,1) &
$\hat{n}^{(2)}$ & (0,1,0)  \\
${\bm r}^{(3)}$   & $\frac{\bar a}{4}$(1,0,1) & $\hat{z}^{(3)}$ & $\frac{1}{\sqrt{3}}$ (-1,1,-1) & 
$\hat{n}^{(3)}$ & (0,0,1) \\
${\bm r}^{(4)}$   & $\frac{\bar a}{4}$(0,1,1) & $\hat{z}^{(4)}$ & $\frac{1}{\sqrt{3}}$ (1,-1,-1)
\end{tabular}
\end{ruledtabular}
\end{table}

%%%%%%%%%%%%%%%%%%%%%%%%%%%%%
\subsection{Mean-field theory}
\label{sec:mft-2}

We are interested in calculating the elastic neutron-scattering cross section 
for both Heisenberg and \Is\ Ising spins on the pyrochlore lattice at the 
mean-field level. 
Therefore, we use the general anisotropic Hamiltonian, $H_H$, as the starting
point for MFT and include 
a local, fictitious field term, $|{\bm h}|=|{\bm h}_i^a|$, 
(where at the end of the calculation ${\bm h}_i^a \rightarrow {\bm 0}$),
\begin{equation}
H_{H} = - \frac{1}{2}\sum_{i,j} \sum_{a,b} \sum_{u,v}
{\mathcal J}^{ab}_{uv}(i,j) S_{i}^{a,u} S_{j}^{b,v}
- \sum_{i,a,u} h_{i}^{a,u} S_{i}^{a,u} ,
\label{eq-HH}
\end{equation}
where
\begin{eqnarray}
\label{eq-JH}
& &{\mathcal J}^{ab}_{uv}(i,j)  =  
J \delta_{R_{ij}^{ab},R_{nn}} \; ({\hat n}^{u} \cdot {\hat n}^{v})  \\ 
& & \ \ \ \ \ \ \ \ + \; \Delta \delta_{i,j} \delta^{a,b} 
\; ({\hat z}^{a}\cdot {\hat n}^{u})({\hat z}^{b}\cdot {\hat n}^{v}) \nonumber \\
& &- D_{dd}  
\left ( \frac{({\hat n}^{u}\cdot {\hat n}^{v})}{|{\bm R}_{ij}^{ab}|^3}
- \frac{3 ({\hat n}^{u}\cdot {\bm R}_{ij}^{ab}) 
({\hat n}^{v}\cdot {\bm R}_{ij}^{ab})}{|{\bm R}_{ij}^{ab}|^5} \right )\; . \nonumber 
\end{eqnarray}
In the notation of this general model, the spin 
vectors are represented by 
${\bm S}_i^a = {\hat n}^{(1)}S_i^{a,1}+{\hat n}^{(2)}S_i^{a,2}+
{\hat n}^{(3)}S_i^{a,3}$, 
where the unit vectors ${\hat n}^{u}$ are the global Cartesian unit 
vectors, see Table \ref{tab-raza}, 
and $S_i^{a,u}$ is the $u$-th component of spin. 
The sum in Eq.~\ref{eq-HH} does not include terms with ${\bm R}_{ij}^{ab}=0$.
For \Is\ Ising spins, one begins with $H_I$, Eq.~\ref{eq-Hising}, and 
adds the field term $-\sum_{i,a} ({\bm h}_i^a \cdot {\hat z}^a) \sigma_i^a$.
The resultant interaction parameter, ${\mathcal J}^{ab}(i,j)$, does not 
include the spin components.   

The general expression for the mean-field free energy is
as follows,
\begin{eqnarray}
\label{eq-F}
F_{\rho} &=& {\rm Tr}\{\rho H_H\} + T {\rm Tr}\{\rho \ln \rho \} \\
&=& \langle H_H \rangle_{\rho} + T \langle \ln \rho \rangle_{\rho} \nonumber,
\end{eqnarray}
where $\rho$ is the many-body density matrix and ${\rm Tr}$ represents a 
trace over the states of $\rho$.
A mean-field form for $F_{\rho}$ is obtained by first expressing  
the many-body density matrix as a product of single-particle density matrices 
$\rho(\{{\bm S}_i^a \}) = \prod_{i,a} \rho_i^a({\bm S}_i^a)$, followed by minimizing
$F_{\rho}$ with respect to $\rho_i^a$ (the variational parameters) 
subject to the constraints ${\rm Tr}\{\rho_i^a \}=1$
and ${\rm Tr}\{\rho_i^a {\bm S}_i^a \}={\bm m}_i^a$, where ${\bm m}_i^a$ 
is the local, vector order parameter. 
For Ising spins, ${\bm m}_i^a$ has only one component, $m_i^a$. 
Next, the resulting mean-field free energy is transformed to momentum space by
applying the definitions, 
\begin{eqnarray}
\label{eq-FTm}
m_{i}^{a,u} = \sum_{\bm q} m_{\bm q}^{a,u} \e^{-\imath {\bm q} \cdot {\bm R}_{i}^{a}}, \\
{\mathcal J}^{ab}_{uv}(i,j) = \frac{1}{N_{\rm cell}}
\sum_{\bm q} {\mathcal J}^{a b}_{uv}({\bm q}) \e^{\imath {\bm q} \cdot {\bm R}_{ij}^{ab}},
\label{eq-FTJ}
\end{eqnarray}
where $N_{\rm cell}$ is the number of fcc Bravais lattice points. 
We note that the above convention for the Fourier transform, which 
employs the position of the spin, ${\bm R}_i^a$, results in 
a real symmetric ${\bm q}$-dependent interaction matrix ${\mathcal J}({\bm q})$,
$12\times 12$ for Heisenberg and $4\times 4$ for Ising spins. 
An alternate convention for the Fourier transform
uses ${\bm R}_i$, the Bravais lattice points, instead of ${\bm R}_i^a$ 
and yields a complex ${\mathcal J}({\bm q})$, refer to 
Appendix~\ref{append:nscatt} for details.
For a non-Bravais lattice, the interaction matrix is not fully 
diagonalized by a Fourier transform. Hence, to completely diagonalize
${\mathcal J}({\bm q})$ one must transform the ${\bm q}$-dependent
variables, ${\bm m}_{\bm q}^a$, to normal mode variables. In component form, 
the normal mode transformation is given by    
\begin{equation}
\label{eq-nmodes}
m_{\bm q}^{a,u} = \sum_{\alpha=1}^{4}\sum_{\mu=1}^{3} U^{a,\alpha}_{u,\mu}({\bm q}) 
\phi_{\bm q}^{\alpha,\mu},
\end{equation}
where the indices ($\alpha,\mu$) label the normal modes 
($12$ for Heisenberg spins), and $\{\phi_{\bm q}^{\alpha,\mu}\}$ 
are the amplitudes of the normal modes. 
In matrix form, $U({\bm q})$ is the unitary matrix that diagonalizes 
${\mathcal J}({\bm q})$ in the spin$\otimes$sublattice 
space with eigenvalues $\lambda({\bm q})$. 
Hence, $U^{a,\alpha}_{u,\mu}({\bm q})$ represents the $(a,u)$ component of
the $(\alpha,\mu)$ eigenvector at ${\bm q}$ with eigenvalue 
$\lambda^{\alpha}_{\mu}({\bm q})$. 
Finally, the mean-field free energy to quadratic order in the normal modes reads, 
\begin{equation}
{\mathcal F}_{\rho}(T) = \frac{1}{2} 
\sum_{{\bm q},\alpha,\mu} (nT - \lambda^{\alpha}_{\mu}({\bm q})) |\phi_{\bm q}^{\alpha,\mu}|^2
- T \sum_{{\bm q},\alpha,\mu} {\tilde h}_{\bm q}^{\alpha,\mu} \phi_{-\bm q}^{\alpha,\mu},
\label{eq-fmf}
\end{equation} 
where ${\mathcal F}_{\rho}(T)=F_{\rho}(T)/N_{\rm cell}$, 
${\tilde h}_{\bm q}^{\alpha,\mu} \propto {\bm h}_{\bm q}^{a,u}/T$,  
$T$ is the temperature in units of $k_{\rm B}$, and
$n=3$ for Heisenberg spins. Note, in order to consider 
the Ising case, the indices $u$ and $\mu$ 
are dropped from Eq.~\ref{eq-fmf} and $n=1$. 
We have also dropped a constant from the expression for ${\mathcal F}_{\rho}(T)$, 
refer to Appendix \ref{append:nscatt}.

The neutron-scattering cross section for unpolarized neutrons in the 
dipole approximation is given by the general expression
\cite{jensen-book,lovesey-book},
\begin{equation}
\label{eq-Xsect1}
\frac{d\sigma(Q)}{d\Omega} = \frac{C [f(Q)]^{2}}{N_{\rm cell}} \sum_{i,j} \sum_{a,b}
\langle {\bm S}_{i\perp}^{a} \cdot {\bm S}_{j\perp}^{b} \rangle
\e^{\imath{\bm Q} \cdot {\bm R}_{ij}^{ab}} \; , \\
\end{equation}
where ${\bm Q}$ is the momentum transfer,
${\bm Q} = {\bm G} + {\bm q}$, ${\bm G}$ is a
reciprocal lattice vector and ${\bm q}$ is a vector in
the first zone, $f(Q)$ is the magnetic form factor of the relevant scattering 
ion, and $C$ is a constant. The spin-spin correlation function only 
involves spin components perpendicular to ${\bm Q}$ (i.e., 
${\bm S}_{i\perp}^{a} = {\bm S}_{i}^{a} - 
({\bm S}_{i}^{a}\cdot {\bm Q}){\bm Q}/|{\bm Q}|^2$)
and can be written as,
\begin{eqnarray} 
\label{eq-corrperp}
\langle {\bm S}_{i\perp}^{a} \cdot {\bm S}_{j\perp}^{b} \rangle &=&
\sum_{u,v} ({\hat n}^{u} \cdot {\hat n}^{v} - ({\hat n}^{u} \cdot {\hat Q})
({\hat n}^{v} \cdot {\hat Q})) \nonumber \\
& & \;\; \times \; \langle S_i^{a,u} S_j^{b,v} \rangle \nonumber \\
&=& \sum_{u,v} \left( \delta^{u,v} - \frac{Q^u Q^v}{|{\bm Q}|^2} \right) 
\langle S_i^{a,u} S_j^{b,v} \rangle \;   
\end{eqnarray}
where ${\hat Q}={\bm Q}/|{\bm Q}|$.
The correlation function $\langle S_i^{a,u} S_j^{b,v} \rangle$
is expressed as a thermal average of the mean-field variables and then
transformed to normal modes, 
\begin{eqnarray}
\label{eq-correl}
\langle S_i^{a,u} S_j^{b,v} \rangle = \sum_{{\bm q},{\bm q}\prime} \sum_{\alpha,\beta} 
\sum_{\mu,\nu} \langle \phi_{\bm q}^{\alpha,\mu} \phi_{{\bm q}\prime}^{\beta,\nu} \rangle 
U^{\alpha,a}_{\mu, u}({\bm q})U^{b,\beta}_{v,\nu}({\bm q}\prime) & &  \\
\: \: \: \times \; \e^{-\imath {\bm q} \cdot {\bm R}_i^a} 
\e^{-\imath {\bm q}\prime \cdot {\bm R}_j^b} \; . \nonumber
\end{eqnarray} 
The correlation function of normal mode variables is calculated from derivatives
of the mean-field partition function, 
$Z={\rm Tr}\{\e^{-\beta {\mathcal F}_{\rho}(T)}\}$, where
${\mathcal F}_{\rho}(T)$ is given by Eq.~\ref{eq-fmf}, with respect to 
${\tilde h}_{\bm q}^{\alpha,\mu}$. 
The result is   
\begin{equation}
\label{eq-correln-H}
\langle \phi_{\bm q}^{\alpha,\mu} \phi_{{\bm q}\prime}^{\beta,\nu} \rangle = 
\frac{\delta^{\alpha,\beta}\delta^{\mu,\nu}\delta_{{\bm q}+{\bm q}\prime,0}}
{(3 - \lambda^{\alpha}_{\mu}({\bm q})/T)}\; ,
\end{equation}  
for Heisenberg spins and
\begin{equation}
\label{eq-correln-I}
\langle \phi_{\bm q}^{\alpha} \phi_{{\bm q}\prime}^{\beta} \rangle = 
\frac{\delta^{\alpha,\beta} \delta_{{\bm q}+{\bm q}\prime,0}}
{(1 - \lambda^{\alpha}({\bm q})/T)} \; ,
\end{equation}
for \Is\ Ising spins.

Using Eqs.~\ref{eq-corrperp}, \ref{eq-correl}, and \ref{eq-correln-H}
or \ref{eq-correln-I} in Eq.~\ref{eq-Xsect1} and carrying out the sums, 
one obtains equations for the scattering cross section. 
In the case of Heisenberg spins, we have
\begin{equation}
\frac{1}{N_{\rm cell}}\frac{d\sigma(Q)}{d\Omega} = C [f(Q)]^{2} \sum_{\alpha,\mu}
\frac{|{\bm F}_{\mu,\perp}^{\alpha}({\bm q})|^2}{(3 - \lambda^{\alpha}_{\mu}({\bm q})/T)},
\label{eq-XsectH}
\end{equation}
where 
\begin{equation}
{\bm F}_{\mu,\perp}^{\alpha}({\bm q})=\sum_{\bm a}
\left\{{\bm U}_{\mu}^{\alpha,a}({\bm q}) - ({\bm U}_{\mu}^{\alpha,a}({\bm q}) \cdot {\hat Q})
{\hat Q} \right\} \e^{\imath {\bm G} \cdot {\bm r}^{a}} 
\label{eq-FperpH}
\end{equation}
is a three-component vector. For \Is\ Ising spins, one has
\begin{equation}
\frac{1}{N_{\rm cell}}\frac{d\sigma(Q)}{d\Omega} = C [f(Q)]^{2} \sum_{\alpha}
\frac{|{\bm F}_{\perp}^{\alpha}({\bm q})|^2}{(1 - \lambda^{\alpha}({\bm q})/T)}\; , 
\label{eq-XsectI}
\end{equation}
with 
\begin{equation}
{\bm F}_{\perp}^{\alpha}({\bm q}) = \sum_{a} \hat{z}_{\perp}^{a} U^{\alpha,a}({\bm q})
\e^{\imath {\bm G} \cdot {\bm r}^{a}} \; ,
\label{eq-FperpI}
\end{equation}
where ${\bm F}_{\perp}^{\alpha}({\bm q})$ is still a three-component vector and 
${\hat z}_{\perp}^a={\hat z}^a - ({\hat z}^a \cdot {\hat Q}) {\hat Q}$. 

Equations \ref{eq-XsectH} and \ref{eq-XsectI} are the main results of this
section. They provide a mean-field description for 
the PM elastic neutron-scattering of Heisenberg and \Is\ Ising moments, respectively, on the 
pyrochlore lattice. The temperature that defines the paramagnetic regime is 
set by the maximum eigenvalue according to 
\begin{equation}
\label{eq-Tc}
T > T_c^{\rm MF} \equiv {\rm max}_{\bm q}\{\lambda^{\rm max}({\bm q})\}/n \ ,
\end{equation} 
where $\lambda^{\rm max}({\bm q})$ is the maximum eigenvalue at wave vector 
${\bm q}$, and ${\rm max}_{\bm q}$ selects the global maximum for all ${\bm q}$. 
The ${\rm max}_{\bm q}\{\lambda^{\rm max}({\bm q})\}$ occurs at the ordering wave
vector ${\bm q}_{\rm ord}$.

%%%%%%%%%%%%%%%%%%%%%%%%%%%%%%%%%%%
\section{Neutron Scattering of \Tb\ \protect}
\label{sec:neutron}

%%%%%%%%%%%%%%%%%%%%%%%%%%%%%%%%%%%%%%%%%%%%

Starting with the zeroth order (low energy) \Is\ description for 
\Tb\,  we use $J_1=-2.64\; {\rm K}$ and $D=0.48\; {\rm K}$  
yielding $J_{\rm nn}/D_{\rm nn}=-1.1$ (Ref.~\onlinecite{tbtio-gingras1}), 
which compares to 
$J_{\rm nn}/D_{\rm nn}=-0.22$ for \Ho\ (Ref.~\onlinecite{hotio-spincorrel}) and 
$J_{\rm nn}/D_{\rm nn}=-0.52$ for \Dy\ (Ref.~\onlinecite{dipSImodel1}).
Therefore, at a nearest-neighbor cutoff distance, 
\Tb\ is an AFM \Is\ pyrochlore that is predicted to 
develop non-collinear AFM order, with
ordering wave vector ${\bm q}_{\rm ord}={\bm 0}$, at $T\approx 1\; {\rm K}$ 
(Refs.~\onlinecite{tbtio-gingras1} and \onlinecite{dipSImodel1}).  
We emphasize that in the context of
a  \Is\ Ising model with $J_{\rm nn}/D_{\rm nn}=-1.1$,
\Tb\  
is still predicted to be a long-ranged AFM when 
both antiferromagnetic nearest-neighbor exchange and long range dipole
interactions are considered \cite{dipSImodel1}. 
Hence, the antiferromagnetic exchange in a \Is\ model of 
\Tb\ is sufficiently strong to prevent the perturbations 
arising from long-range dipolar interactions  
from changing the ordered state of the model.
The counter point to the above model predictions is that experimentally 
\Tb\ remains a collective paramagnetic down to very low temperatures, 
$T \stackrel{>}{\sim} 50\; {\rm mK}$ (Ref.~\onlinecite{gardner-tbtio50}).

\begin{figure}[!ht]
\begin{center}
\includegraphics[width=3.0in,height=7.0in]{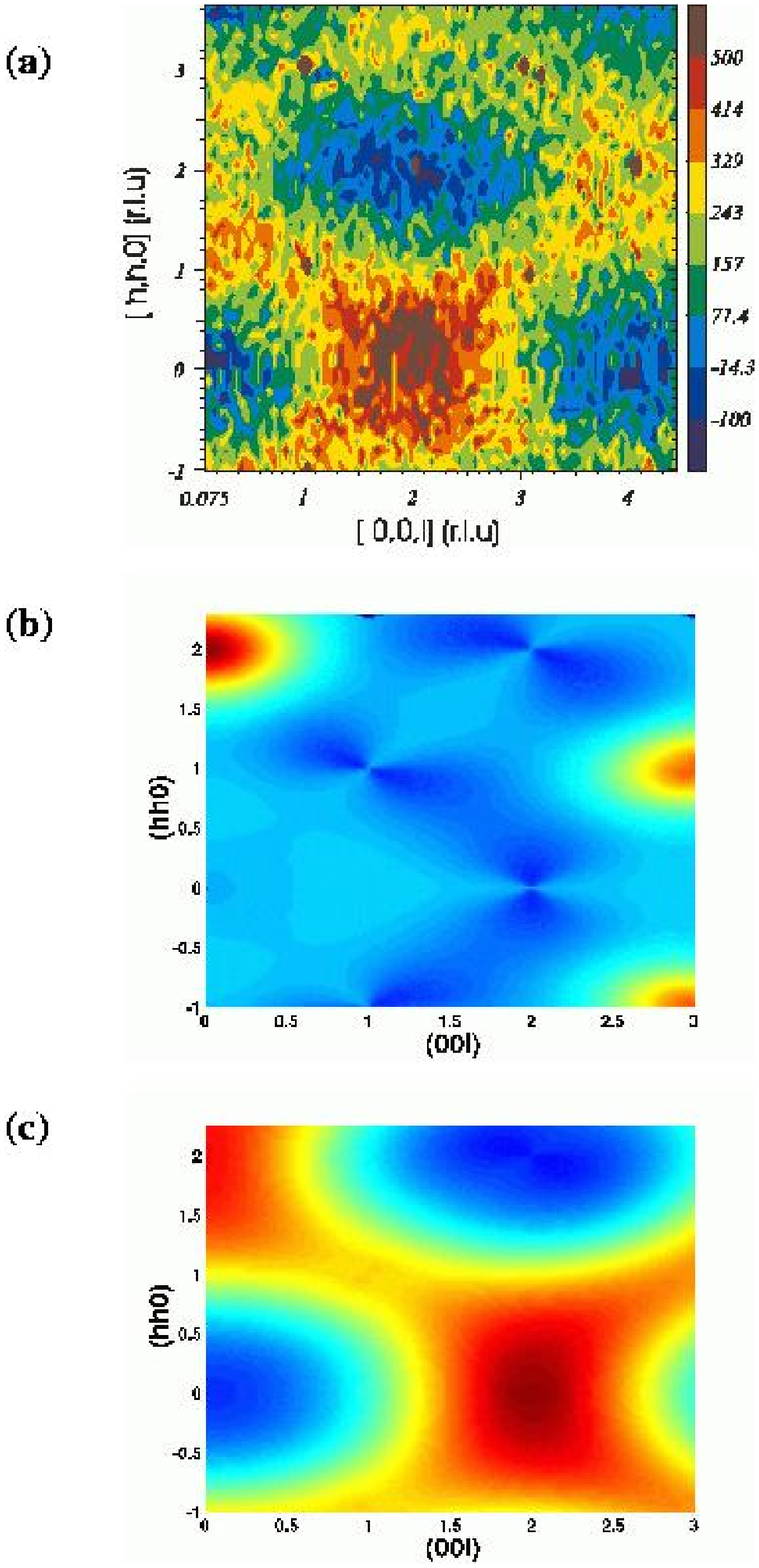}
\caption{(Color online) Paramagnetic scattering in the $(hhl)$ plane for \Tb\ :
(a) Experimental paramagnetic scattering \cite{tbtio-gardner2}$(T=9\, {\rm K})$,
maximum intensity at ${\bm Q}=0,0,2$, 
(b) MF model of \Tb\ treated as a \Is\ Ising pyrochlore  $(T=1.5T_c^{\rm MF})$,
no intensity at ${\bm Q}=0,0,2$,
(c) MF model of \Tb\ treated as an anisotropic Heisenberg pyrochlore
$(T=1.5T_c^{\rm MF}, \Delta=20\, {\rm K})$, maximum intensity at ${\bm Q}=0,0,2$.}
\label{fig-nsTb}
\end{center}
\end{figure}

As an initial attempt to explain the physics of \Tb , we
investigate the PM correlations within MFT and compare 
to the experimental results for elastic neutron-scattering. 
Experimental data for elastic neutron-scattering in \Tb\ are shown 
in Fig.~\ref{fig-nsTb}(a). The most intense
region of scattering is centered around ${\bm Q}=0,0,2$ with reduced 
correlations extending toward 
${\bm Q}=2,2,0$ 
 and a scattering minimum
at ${\bm Q}={\bm 0}$. From the pyrochlore 
lattice structure and the MF formalism, we know 
that the intensity at ${\bm Q}=0,0,2$ is controlled
by the eigenmodes at ${\bm Q}=0,0,0$ but modulated by the phase factor 
$\exp(\imath {\bm G} \cdot {\bm r}^a)$, see Eqs.~\ref{eq-XsectI} and
~\ref{eq-FperpI}.
This raises the question as to whether the maximum about ${\bm Q}=0,0,2$ 
could be interpreted as the precursor of a long-range ordered 
non-collinear AFM state.

We begin by considering \Is\ Ising spins on the pyrochlore lattice.  
The details upon which our
arguments here are based are provided in Appendix \ref{append:symm}.
The neutron-scattering intensity profile 
is determined by ${\bm F}_{\perp}^{\alpha}({\bm q})$, Eq.~\ref{eq-FperpI}, 
and contains information on the spin anisotropy via ${\hat z}^a$ and 
the eigenvalues $\lambda^{\alpha}_{\mu}({\bm q})$, 
and the symmetry of the lattice through the eigenvectors, 
$U^{\alpha,a}({\bm q})$,
and a phase factor, $\exp{(\imath {\bm G} \cdot {\bm r}^{a})}$. Hence,
the nature and strength of the exchange and dipole-dipole
interactions are arbitrary. 
From these basic symmetry components, we find that 
$|{\bm F}_{\perp}^{\alpha}(0,0,0)|^2=|{\bm F}_{\perp}^{\alpha}(0,0,2)|^2$,
or that the scattering intensity about ${\bm Q}=0,0,0$ and ${\bm Q}=0,0,2$ has
the same numerical value, disregarding the form factor ($f(Q)$).
An equivalent statement is the intensities
about ${\bm Q}=0,0,0$ and ${\bm Q}=0,0,2$ are symmetry related.  This strong
condition on the scattering pattern is in serious contradiction with the
experimentally observed results.  
In contrast, if we consider an anisotropic Heisenberg pyrochlore model 
(Eq.~\ref{eq-XsectH} with finite $\Delta$), we find that the lattice and 
spin degrees of freedom do not force the scattering intensity to be 
identical about ${\bm Q}=0,0,0$ and ${\bm Q}=0,0,2$. 
For a model with full $O(3)$ spin symmetry, 
the scattering profile is controlled by 
${\bm F}_{\mu, \perp}^{\alpha}({\bm q})$, Eq.~\ref{eq-FperpH}. 
The significant difference between Eqs.~\ref{eq-FperpH} and
\ref{eq-FperpI} is the restoration of spin isotropy, i.e., 
the geometric factor defining the local \Is\ quantization 
axis, ${\hat z}^a$, is absent from Eq.~\ref{eq-FperpH}.
Therefore, on purely symmetry grounds, 
no \Is\ Ising model 
(i.e., Hamiltonian) with arbitrary distance dependent $J_{ij}({\bf r_{ij}})$
 for \Tb\ will reproduce the 
experimental PM correlations
shown in Fig.~\ref{fig-nsTb}(a). 
Earlier works have recognized the need to consider more isotropic spin models
for \Tb . In Ref.~\onlinecite{tbtio-gardner2}, the qualitative features of the 
PM scattering were reproduced from an isotropic structure factor
for the nearest-neighbor Heisenberg pyrochlore AFM.   
Similar results were obtained in Ref.~\onlinecite{tbtio-yasui1} 
by considering specific ${\bm Q}={\bm 0}$ 
spin structures on a cluster of two tetrahedra.

\begin{figure}[!ht]
\begin{center}
\includegraphics[width=3.0in,height=2.5in]{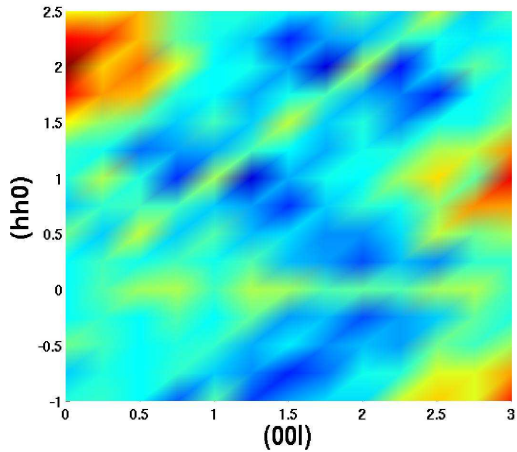}
\caption{(Color online) Monte Carlo results for paramagnetic 
scattering in the $(hhl)$ plane for \Tb\
as a \Is\ Ising AFM, $T=5.0$K, $N=1024$ spins. Note that there is no intensity about
${\bm Q}=0,0,2$. Dipoles were treated via the Ewald method.}
\label{fig-nsTb-mc}
\end{center}
\end{figure}

To support the picture obtained on symmetry grounds, 
we have applied our MF formalism to two models for \Tb : $(1)$ a 
pyrochlore system with \Is\ Ising spins, and 
$(2)$ a pyrochlore lattice with Heisenberg spins and finite 
\Is\ anisotropy. In both cases, the dipole-dipole interactions are
evaluated with the Ewald method, see Appendix \ref{append:ewald}.
For the \Is\ Ising description 
(i.e., $\Delta/|J| \gg 1$ and $\Delta/D \gg 1$ and Eq.~\ref{eq-XsectI}), 
we use $J_1=-2.64\, {\rm K}$ and $D=0.48\, {\rm K}$ (Ref.~\onlinecite{tbtio-gingras1}).
Our data are shown in Fig.~\ref{fig-nsTb}(b). 
Note that the scattering about ${\bm Q}=0,0,0$ and ${\bm Q}=0,0,2$ is 
symmetry related, as predicted above, but is an intensity minimum. 
Monte Carlo data for this \Is\ Ising model agrees 
with our MF results, see Fig.~\ref{fig-nsTb-mc}.  
Monte Carlo simulations for \Tb\ as a \Is\ Ising dipolar model were
performed on a $L=4$ lattice ($N=1024$ spins) at $T=5$ K ($T_c^{\rm MF} \sim 1$ K)
with $J_1=-2.64 K$ and $D=0.48$ K, with the dipolar sum treated via the
Ewald summation method.
neutron-scattering data (as determined by Eq.~\ref{eq-Xsect1} 
(Ref.~\onlinecite{hotio-spincorrel})) were collected
after 5$\times 10^7$ Monte Carlo steps per spin for both equilibration
and measurement stages, and are shown in Fig.~\ref{fig-nsTb-mc}.
The intensity minimum at ${\bm Q}=0,0,0$ and ${\bm Q}=0,0,2$ supports the above
mean-field results and symmetry arguments.
For a Heisenberg model with finite anisotropy  
(i.e., $\Delta/|J| > 1$ and $\Delta/D > 1$ and Eq.~\ref{eq-XsectH}), 
our MF results are provided in Fig~\ref{fig-nsTb}(c). 
With an anisotropy strength of $\Delta=20$K 
(i.e., $\Delta/D \approx 41.7$), we achieve good 
qualitative agreement with the experiment. The region around
${\bm Q}=0,0,2$ has the strongest scattering with reduced intensity
near ${\bm Q}=2,2,0$ and the interconnecting regions. If we 
turn off the finite anisotropy ($\Delta=0$), i.e.,  
an isotropic Heisenberg model with long-range dipoles, 
the dominating scattering remains about points ${\bm Q}=0,0,2$ and ${\bm Q}=2,2,0$,   
but there is increased intensity along the bridge regions  
in ${\bm q}$-space connecting these points. 
Finally, in the absence of dipoles and $\Delta=0$ one has the 
near neighbor AFM exchange Heisenberg model, where the scattering
intensity forms a network of interconnected triangles with 
equal intensity about ${\bm Q}=0,0,2$ and ${\bm Q}=2,2,0$ 
(Refs.~\onlinecite{tbtio-gardner2,moessner-chalker1,canals-cjp}). 
Hence, a partial restoration of the spin isotropy is sufficient to 
place scattering about points ${\bm Q}=0,0,2$ and ${\bm Q}=2,2,0$ in 
${\bm q}$-space, but to achieve good qualitative agreement with
the experimental intensity profile dipolar interactions are 
necessary as is a finite single-ion contribution to the 
Hamiltonian.   

%%%%%%%%%%%%%%%%%%%%%%%%%%%%%%%%%%%%%%%%%%%%%%%%%%%%%%
\section{Discussion \protect}
\label{sec:disc}

\subsection{Tb$_2$Ti$_2$O$_7$}

The prediction of an \Is\ Ising model for \Tb\ is that of an AFM long-range ordered 
state in which all spins point either in or out of the unit tetrahedra
at $T\approx 10^0\, {\rm K}$. However, this prediction is not realized 
experimentally. Another problem, 
and possibly even more important, with a \Is\ Ising model for 
\Tb , is that the mean-field 
PM correlations, see Fig.~\ref{fig-nsTb}(b), do not
agree with the experimentally observed results, Fig.~\ref{fig-nsTb}(a). 
In Section \ref{sec:neutron} and in Appendix \ref{append:symm}, we
demonstrated on symmetry grounds alone that no pyrochlore model 
with \Is\ Ising moments could reproduce the basic features of 
the experimental PM scattering, i.e., strongest 
intensity centered about ${\bm Q}=0,0,2$, lower intensity
about ${\bm Q}=2,2,0$, and minimum intensity 
at the zone center ${\bm Q}=0,0,0$.
These symmetry arguments were also supported by MF and MC 
calculations of the elastic neutron-scattering cross section, 
Eq.~\ref{eq-XsectI}.

In Section \ref{sec:neutron} and in Appendix \ref{append:symm}, 
we were also able to demonstrate on symmetry grounds that  
a Heisenberg pyrochlore model of \Tb\ 
would allow for elastic scattering about
${\bm Q}=0,0,2$ while at the same time permit no scattering about ${\bm Q}=0,0,0$. 
A MF calculation of the PM neutron-scattering in the 
$(hhl)$ plane for an anisotropic Heisenberg model with 
long-range dipoles, Eq.~\ref{eq-Hheis}, 
shows good agreement with the experimental
scattering pattern (strongest intensity about ${\bm Q}=0,0,2$ with
reduced scattering at ${\bm Q}=2,2,0$ and along the bridges between 
these points, see Fig.~\ref{fig-nsTb}(c)).  
Both the finite anisotropy ($\Delta/|J| > 1$ and $\Delta/D > 1$) 
and long-range dipoles are necessary to achieve a quantitative
match with the experiments. Reducing either the single-ion anisotropy 
to the isotropic limit ($\Delta = 0$) or the range of the
dipole-dipole interactions in the model reduces this agreement 
by altering the ratio of scattering intensity 
between ${\bm Q}=0,0,2$ and ${\bm Q}=2,2,0$. However, even a 
dramatically simplified model which would have 
dipoles cutoff at the first nearest-neighbor, $r_c=1$, 
would still improve the picture provided by a nearest-neighbor AFM 
exchange only Heisenberg model presented in Ref.~\onlinecite{tbtio-gardner2}. 
Therefore, and foremost in our argument,
a restoration of the spin isotropy is absolutely 
necessary to place paramagnetic scattering about the ${\bm q}$-space points
${\bm Q}=0,0,2$ and ${\bm Q}=2,2,0$. The dipolar interactions are then 
important for shifting (i.e., redistributing) 
the scattering intensity from ${\bm Q}=2,2,0$ to ${\bm Q}=0,0,2$. 
Intermediate regions between these two points also 
experience a reduction in scattering. In terms of the underlying soft-mode 
spectrum, the ${\bm q}={\bm 0}$ eigenvalues and eigenvectors control 
the scattering at ${\bm Q}=2,2,0$ to ${\bm Q}=0,0,2$. A shift in intensity from
${\bm Q}=2,2,0$ to ${\bm Q}=0,0,2$ signals a PM spin structure that prefers to lie 
in $xy$-plane (i.e., neutron-scattering at ${\bm Q}=0,0,2$ comes from spins with
components perpendicular to this direction). 

Switching to a model with fully isotropic Heisenberg spins 
(as in Ref.~\onlinecite{tbtio-gardner2})
restores all the spin symmetry in the paramagnetic limit.
This picture is dramatically inconsistent with the experimentally 
determined single-ion structure of Tb$^{3+}$ 
(${\rm J}=6$, $^{7}{\rm F}_{6}$) in \Tb\ , 
where a ground state doublet is separated from the first excited
doublet by an anisotropy gap of $20\, {\rm K}$, close to the $\theta_{\rm CW}$ 
temperature \cite{hotio-rosenkranz,tbtio-gingras1}. Therefore, a restoration of the
full spin symmetry at $T<20\, {\rm K}$ seems an unlikely explanation for 
the PM scattering at $9\, {\rm K}$.

The current MF approach does not allow the single-ion properties to be 
systematically considered, but a RPA calculation does \cite{tbtio-yjkao}. 
By retaining only the 
simplest energy level structure in the Tb$^{3+}$ wave function, 
the ground state doublet and the first excited state doublet, 
one can relax the strict \Is\ Ising constraint on the spins in
a controlled approximation. 
Within the RPA, fluctuations out of the ground state and into the 
first excited state levels are equivalent to a fluctuating canting of spins 
away from the Ising geometry. In this case, the 
lowest order fluctuations from the strict \Is\ Ising limit 
yield qualitative agreement with experiment for the paramagnetic spin-spin correlations 
\cite{tbtio-yjkao}. Others have also proposed a simple relaxation scheme of the
strict \Is\ Ising directions \cite{tbtio-yasui1}. 
Theoretical work remains to be done to explain the failure of \Tb\ to order at 
a temperature of $1\, {\rm K}$, and why it remains paramagnetic down to 
$50$ mK (Refs.~\onlinecite{tbtio-gardner1} and \onlinecite{tbtio-gardner2}).

%%%%%%%%%%

\subsection{General Discussion: Avenues for Other Studies}

We now briefly discuss some puzzling experimental results for a 
few highly frustrated magnets. We note that the present mean-field formulation
for the structure factor, $S({\bm q})$, could provide valuable insight on the 
development of magnetic correlations out of the PM regime for each of these 
systems. The first, very paradoxical, system is the antiferromagnetic 
Gd$_3$Ga$_5$O$_{12}$ garnet (GGG).
This material, where Gd$^{3+}$ is the magnetic ion with a spin $S=7/2$,
consists of two sublattices of intertwined spirals of corner-sharing triangles.
For classical Heisenberg spins coupled by nearest-neighbor antiferromagnetic 
exchange, each spiral on a garnet lattice structure 
should display a thermally-induced spin-nematic order-by-disorder 
transition according to work by
Moessner and Chalker \cite{moessner-chalker1}. 
 Some precursors of spin coplanarity in GGG may have recently been observed in M\"ossbauer
experiments \cite{bonville-ggg}.
In GGG, however,
dipolar interactions are approximately 50\% of the strength of the exchange interactions
for nearest-neighbors and are, consequently, a sizable perturbation to contend with in this 
system \cite{ggg-wolf}.
In zero applied magnetic field, specific heat, magnetic susceptibility 
and nonlinear susceptibility measurements 
on GGG strongly suggest that this material undergoes a spin-glass transition around 
$140$ mK (Refs.~\onlinecite{ggg-shiffer} and \onlinecite{ggg-shiffer2}). However, 
the nonlinear susceptibility, $\chi_{\rm nl}$, 
measurements indicate that the spin-glass transition in this material is unusual in that 
$\chi_{\rm nl}$ exhibits two maxima\cite{ggg-shiffer2}.
In contrast to bulk measurements\cite{ggg-shiffer,ggg-shiffer2},
 neutron-scattering experiments on powder samples of isotopically enriched 
$^{160}$Gd (natural Gd has a huge neutron absorption cross section) indicates the 
development of spin-spin correlations at approximatively 
$140$ mK, extending to a length scale of $\sim 100\AA$ 
(Refs.~\onlinecite{ggg-pentrenko} and \onlinecite{ggg-pentrenko-prl}). 
It is unclear at present whether or not
the development of extended spin correlations in GGG at 
$\sim 140$ mK is an intrinsic effect
or is due to material impurities and/or defects (e.g.,
Gd$^{3+}$ magnetic ions at Ga$^{3+}$ sites \cite{ramirez-private}).
Another interesting system is the Gd$_2$Ti$_2$O$_7$ 
pyrochlore antiferromagnet, where
Gd$^{3+}$ is also the moment-carrying species. In Gd$_2$Ti$_2$O$_7$, the 
dipolar interactions are approximately $20\%$ of the strength of the exchange interactions
for nearest-neighbors and is here, just as in GGG above, 
an important perturbation \cite{gdtio-raju,gdtio-palmerchalker,gdtio-ramirez}.
Palmer and Chalker argue that the ground state consists of a 
fully ordered structure where each tetrahedral unit cell has an identical (zero 
total magnetic moment) spin configuration 
(a so-called ${\bm q_{\rm ord}}={\bm 0}$ structure)
\cite{gdtio-palmerchalker}.
Recent work has confirmed that this ground state is extremely robust against quantum
fluctuations \cite{delMaestro}.
However, recent experiments on Gd$_2$Ti$_2$O$_7$ are rather puzzling and 
appear inconsistent with Palmer and Chalker's work. Specifically, 
neutron diffraction measurements on $^{160}$Gd isotopically enriched
powders find a partially ordered phase with one
disordered sublattice and propagation vector 
${\bm q}=\frac{1}{2}, \frac{1}{2}, \frac{1}{2}$ at 
$T=50$ mK (Ref.~\onlinecite{gdtio-bramwell}), hence incompatible with
the predictions of Palmer and Chalker \cite{gdtio-palmerchalker}.
Specific heat measurements find strong evidence for two transitions at
$T=0.7$ K and at $T=1.0$ K (Ref.~\onlinecite{gdtio-ramirez}). 
Recent mean-field calculations find evidence for a two-step magnetic
ordering in this system.\cite{enjalran-gdtio,cepas}
So for Gd$_2$Ti$_2$O$_7$, there also exists a complex behavior 
as signaled by thermodynamic measurements, 
theoretical predictions and neutron-scattering results.~\cite{bramwell-gto-JPC}
Finally, the Yb$_2$Ti$_2$O$_7$ pyrochlore ferromagnet is also puzzling
\cite{ybtio-hodges,yasui-yto}. There, neutron-scattering results, muon spin relaxation  
and M\"ossbauer experiments 
suggest a ground-state that lacks long-range magnetic order while there exist
good evidence from the M\"ossbauer data that a first order spin freezing transition 
occurs around $T_f\sim 0.24$ K. 
Meanwhile, elastic neutron-scattering results
reveal the development of nontrivial spin-spin correlations as 
$T_f$ is approached from above.
In this system too, it is possible that long-range dipolar
interactions may play some role due to the contribution of the 
$\vert J, m_J\rangle = \vert 7/2, \pm 7/2\rangle$ eigenstate within 
the effective $S_{\rm eff}=1/2$ ground state doublet.~\cite{ybtio-hodges} 

%%%%%%%%%%%%%%%%%%%%%%%%

%%%%%%%%%%%%%%%%%%%%%%%%%%%%%%%%%%%%%%%%%%%%%%%%%%%%%%
\section{Conclusions \protect}
\label{sec:concl}

In conclusion, we 
have demonstrated on symmetry grounds, through 
MF calculations, and MC simulations that the experimentally measured PM 
elastic neutron-scattering in \Tb\ is inconsistent with a \Is\ Ising pyrochlore 
spin structure. From the qualitative agreement obtained using an 
anisotropic Heisenberg model, we argue in favor of a more isotropic 
effective spin model to describe the low energy phases of \Tb .     

Finally, we have discussed the usefulness of a combined mean-field theory  
and Ewald method approach to studying geometrically frustrated magnets with long-range
dipole-dipole interactions in the paramagnetic phase. This approach could be 
applied in general to any geometrically frustrated system. 
The zero field picture of Gd$_3$Ga$_5$O$_{12}$ (GGG) is particularly interesting 
because the low temperature phase remains an unresolved issue that
entails unraveling the physics of competing exchange and long-range dipole-dipole 
interactions in a garnet lattice environment
\cite{ggg-shiffer,ggg-shiffer2,ggg-pentrenko,ggg-pentrenko-prl,ggg-dunsiger,bonville-ggg}. 

%%%%%%%%%%%%%%%%%%%%%%%%%%%%%%%%%%%%%%%%%%%%%%%%%%%%%%%%
\begin{acknowledgments}
We would like to acknowledge Bill Buyers, Steve 
Bramwell, Adrian del Maestro, Jason Gardner, Bruce Gaulin, 
Byron den Hertog, Ying-Jer Kao,
Jean-Yves Delannoy, Roger Melko, Adrian del Maestro, 
Hamid Molavian, and Taras Yavor\'{s} kii for many useful discussions. 
We would also like to thank Adrian del Maestro for a 
careful reading of the manuscript. 
This work is supported by NSERC of Canada, 
the Canada Research Chairs Program, Research Corporation and the Province of Ontario.
\end{acknowledgments}

\appendix
%%%%%%%%%%%%%%%%%%%%%%%%%%%%%%%%%%%%%%%%%%%%%%%%%%%%%%%%%%%%%%%%
\section{Neutron scattering in the Gaussian approximation \protect}
\label{append:nscatt}

As we noted in the Introduction, other authors have discussed MFT
and its application to 
magnetism \cite{chaikin-lubensky,abharris-mft,reimers-mft,rbs-mft,kanda-mft-Ho}
and the Ewald method for magnetic dipoles \cite{CK,AF,litbf-holmes}.
Our purpose is to combine the techniques of MFT, developed here, with the 
and Ewald procedures for magnetic dipoles in ${\bm q}$-space, developed in 
Appendix \ref{append:ewald}, so they can be readily applied to other problems
of highly frustrated rare-earth magnets.

In this appendix we provide a detailed derivation of the mean-field 
equations for the elastic scattering cross section for pyrochlore
spin systems.  
Our derivation is performed for Heisenberg spins, with 
a finite local \Is\ anisotropy, in order to broaden the appeal of the results. 
Connections to \Is\ Ising systems, infinite local \Is\ anisotropy, 
are noted at appropriate points.   
As mentioned in Section \ref{sec:mft}, the MFT is developed
via a variational approach (VMFT), and in general this approach   
applies to a large array of statistical models with 
arbitrarily complex order parameters \cite{chaikin-lubensky}. 
In this work, the method reproduces the Gaussian approximation 
(GA) of the Landau free energy.
The VMFT described here has been used by others 
\cite{abharris-mft,rbs-mft,reimers-mft,kanda-mft-Ho}, and 
we provide a detailed presentation here 
to clear up the notational inconsistencies that appeared in 
some of these previous works. We also wish to 
provide for comparison with the RPA, 
which allows for a more controlled relaxation of the \Is\ Ising 
restriction \cite{tbtio-yjkao,jensen-book} that experimental evidence
suggests is needed for \Tb\ \cite{tbtio-gardner2,tbtio-yasui1}. 
 
We begin with the model Hamiltonian of Eq.~\ref{eq-Hheis}, $H_H$.  
The conventions for the spin vectors, ${\bm S}_{i}^{a}$, 
unit vectors, ${\hat n}^{u}$ and ${\hat z}^{a}$, and the
description of the pyrochlore lattice in a rhombohedral basis with
a four atom unit cell are as described in Section \ref{sec:mft}. 
Therefore, our starting Hamiltonian for MFT is given by,
\begin{equation}
H_H = - \frac{1}{2}\sum_{i,j} \sum_{a,b} \sum_{u,v}
{\mathcal J}^{ab}_{uv}(i,j) S_{i}^{a,u} S_{j}^{b,v}
- \sum_{i,a,u} h_{i}^{a,u} S_{i}^{a,u} ,
\label{eq-Hgen}
\end{equation}
where ${\mathcal J}^{ab}_{uv}(i,j)$ is defined by Eq.~\ref{eq-JH}, and
a fictitious field term has been added. The field term, with 
$|{\bm h}_i^a|=|{\bm h}|$, is removed from the final equations 
by taking ${\bm h}_i^a \rightarrow 0$. We note that indices
$(a,b)=1,2,3,4$ label the sublattices and $(u,v)=1,2,3$ label the spin 
components.  

In VMFT, an approximate free energy as a function of a trial 
density matrix, ${\rho}$, is formed,
\begin{equation}
F_{\rho} = {\rm Tr}\{\rho H_H \} + T {\rm Tr}\{\rho \ln \rho \} = 
\langle H_H \rangle_{\rho} + T \langle \ln \rho \rangle_{\rho}, 
\label{eq-freeE1} 
\end{equation}
where ${\rm Tr}$ represents a trace over spin variables. 
$F_{\rho}$ is variational and defines an upper bound
to the actual free energy, i.e., $F_{\rho} \ge F$. 
The best functional from for trial density matrix is 
obtained by minimizing $F_{\rho}$ with respect to the 
parameters of $\rho$. For a system of $N$ particles, the MF form of the 
$N$-body density matrix is given by a product of single particle density matrices,
\begin{equation}
\rho(\{{\bm S}_i^a \}) = \prod_{i,a} \rho_{i}^{a}({\bm S}_{i}^{a}) \ .
\label{eq-rhomft} 
\end{equation}
The single particle density matrix, $\rho_{i}^{a}$, is treated as a variational
parameter that is subject to the constraints
\begin{eqnarray}
\label{eq-constraint1}
{\rm Tr}\{\rho_{i}^{a}\} = 1 \ , \\
{\rm Tr}\{\rho_{i}^{a}{\bm S}_{i}^{a}\} = {\bm m}_{i}^{a} \ ,
\label{eq-constraint2}
\end{eqnarray}
which keep the internal energy constant, i.e., $Tr\{\rho H\}=C$. Here,
${\bm m}_{i}^{a}$ is a vector order parameter, for \Is\ Ising spins 
one has the scalar equivalent, ${\rm Tr}\{\rho_{i}^{a}\sigma_{i}^{a}\}=m_{i}^{a}$. 
Incorporating the constraints
into the expression for $F_{\rho}$ gives,
\begin{eqnarray}
F_{\rho} &=& {\rm Tr}\{\rho H_H \} + T\ {\rm Tr}\{\rho \ln \rho \} \\
&-& T\ {\rm Tr}\{\sum_{i,a} \xi_i^a (\rho_i^a - 1) \} \nonumber \\
&-& T\ {\rm Tr}\{\sum_{i,a}(\rho_i^a {\bm S}_i^a - {\bm m}_i^a) \cdot {\bm A}_i^a \}
\nonumber \ ,
\label{eq-freeE2} 
\end{eqnarray}
where $\xi_i^a$ and ${\bm A}_i^a$ are the Lagrange multipliers for the constraints 
of Eq.~\ref{eq-constraint1} and Eq.~\ref{eq-constraint2}, respectively.
In minimizing $F_{\rho}$ with respect to $\rho_i^a$, one has the following results,
\begin{eqnarray}
\frac{\delta}{\delta \rho_i^a} {\rm Tr}\{\rho H_H \} &=& 0 \nonumber \ , 
\\ 
\frac{\delta}{\delta \rho_i^a} {\rm Tr}\{\rho \ln \rho \} &=& 
{\rm Tr}\{ \ln \rho_i^a \} + {\rm Tr}\{ 1 \} \nonumber \ ,  
\\
\frac{\delta}{\delta \rho_i^a} {\rm Tr}\{\sum_{i,a} \xi_i^a \rho_i^a \} &=& 
{\rm Tr}\{\xi_i^a \} \nonumber \ ,
\\
\frac{\delta}{\delta \rho_i^a} T\ {\rm Tr}\{\sum_{i,a} \rho_i^a {\bm S}_i^a \cdot{\bm A}_i^a \}
&=& {\rm Tr}\{ {\bm S}_i^a \cdot {\bm A}_i^a \} \nonumber \ .
\end{eqnarray}
The optimum form for the density matrix is found by solving 
$\delta F_{\rho}/\delta \rho_i^a = 0$,
\begin{equation}
\label{eq-rho1} 
\rho_i^a = {\mathcal C}_i^a \e^{{\bm A}_i^a \cdot {\bm S}_i^a} \ ,
\end{equation}
where ${\mathcal C}_i^a = \e^{\xi_i^a - 1} = 1/{\rm Tr}\{\e^{{\bm A}_i^a \cdot {\bm S}_i^a}\}$
follows from Eq.~\ref{eq-constraint1}. 
Evaluating the trace in ${\mathcal C}_i^a$ for both Heisenberg and Ising spins we 
obtain  
\begin{eqnarray}
\label{eq-ZH}
{\mathcal Z}_i^a &=& \frac{(2\pi)^{3/2}}{(|{\bm A}_i^a|)^{1/2}} I_{1/2}(|{\bm A}_i^a|) \\
&=& \frac{4\pi}{|{\bm A}_i^a|} \sinh(|{\bm A}_i^a|) \nonumber \; , 
\end{eqnarray}
and 
\begin{equation}
\label{eq-ZI}
{\mathcal Z}_i^a = 2\cosh(A_i^a) \ ,
\end{equation}
respectively, where $I_{1/2}(|{\bm A}_i^a|)$ is a modified Bessel function.
The variational local density matrix is now written as
\begin{equation}
\label{eq-rho2} 
\rho_i^a = \frac{\e^{{\bm A}_i^a \cdot {\bm S}_i^a}}{{\mathcal Z}_i^a} 
\end{equation}
and is used, along with the constraints of Eqs.~\ref{eq-constraint1}
and \ref{eq-constraint2}, to rewrite the variational free energy, 
\begin{eqnarray}
\label{eq-FEmft1}
F_{\rho} &=& - \frac{1}{2}\sum_{i,j}\sum_{a,b}\sum_{u,v}
{\mathcal J}^{ab}_{uv}(i,j) m_{i}^{a,u} m_{j}^{b,v}  \\
&-& \sum_{i,a,u} h_{i}^{a,u} m_{i}^{a,u}
+ T \sum_{i,a}\left({\bm A}_i^a \cdot {\bm m}_i^a - \ln {\mathcal Z}_{i}^{a}\right). \nonumber
\end{eqnarray} 

We want an expression for the free energy to quadratic order in 
${\bm m}_i^a$. This means
that one must expand $\ln {\mathcal Z}_i^a({\bm A}_i^a)$ and 
then express ${\bm A}_i^a$ as
a function of ${\bm m}_i^a$. From the series representation of 
${\mathcal Z}_i^a({\bm A}_i^a)$ followed by the series expansion 
of $\ln(1-x)$, one has  
\begin{equation}
\label{eq-expandZ}
\ln {\mathcal Z}_i^a \approx \ln C_1 + \frac{|{\bm A}_i^a|^2}{2n} \ ,
\end{equation}  
where $C_1$ is a model dependent constant and $n=1,3$ for Ising and Heisenberg
spins, respectively. Using Eq.~\ref{eq-constraint2}, one obtains the expression
\begin{eqnarray}
\label{eq-mAH}
{\bm m}_i^a &=& {\hat A}_i^a \frac{I_{3/2}(|{\bm A}_i^a|^2)}{I_{1/2}(|{\bm A}_i^a|^2)} \\
&=& {\hat A}_i^a \left \{ \coth(|{\bm A}_i^a|) - \frac{1}{|{\bm A}_i^a|} \right\}
\end{eqnarray}
for Heisenberg
and
\begin{equation}
\label{eq-mAI}
m_i^a = \tanh (A_i^a)
\end{equation}
for Ising spins.
To first order, we have
\begin{equation}
\label{eq-mAH2}
{\bm m}_i^a = 3{\bm A}_i^a \ , 
\end{equation}
and
\begin{equation}
\label{eq-mAI2}
m_i^a = A_i^a \ . 
\end{equation}
Using Eqs.~\ref{eq-expandZ}-\ref{eq-mAI2}, we can write the MF free energy
to quadratic order in the order parameter, 
\begin{eqnarray}
F_{\rho} &=& \frac{1}{2}\sum_{i,j} \sum_{a,b} \sum_{u,v}
m_{i}^{a,u} \left\{ nT\delta_{i,j}\delta^{a,b}\delta^{u,v}
- {\mathcal J}^{ab}_{uv}(i,j) \right\} m_{j}^{b,v} \nonumber \\
&-& \sum_{i,a,u} h_{i}^{a,u} m_{i}^{a,u} - TpN_{\rm cell} \ln C_1 \ ,
\label{eq-FEreal}
\end{eqnarray}
where $p=4$ denotes the size of the basis (sublattice). As a side note, the 
Lagrange multiplier ${\bm A}_i^a$ can be interpreted as
an effective mean-field interacting with a local moment. Minimizing $F_{\rho}$,
of Eq.~\ref{eq-FEmft1}, with respect to $m_i^{a,u}$ one has
\begin{equation}
A_i^{a,u} = \frac{{\bar h}_i^{a,u}}{T} \ ,
\end{equation}
where 
\begin{equation}
{\bar h}_i^{a,u} = \sum_{j,b,v} {\mathcal J}^{ab}_{uv}(i,j) m_j^{b,v} + h_i^{a,u} \ , 
\end{equation}
is the $u$-th component of the effective field at site $(i,a)$. 
We next exploit the fact that the pyrochlore lattice has the underlying symmetry of
a fcc lattice by defining the Fourier transforms, 
\begin{eqnarray}
\label{eq-ftm}
m_{i}^{a,u} = \sum_{\bm q} m_{\bm q}^{a,u} \e^{-\imath {\bm q} \cdot {\bm R}_{i}^{a}} \ , \\
{\mathcal J}^{ab}_{uv}(i,j) = \frac{1}{N_{\rm cell}}
\sum_{\bm q} {\mathcal J}^{a b}_{uv}({\bm q}) \e^{\imath {\bm q} \cdot {\bm R}_{ij}^{ab}} \ ,
\label{eq-ftj}
\end{eqnarray} 
where $N_{\rm cell}$ is the number of fcc Bravais lattice points, ${\bm R}_i^a$ denotes
the position of a spin, and ${\bm R}_{ij}^{ab} = {\bm R}_i^a - {\bm R}_j^b$. 
Equations \ref{eq-ftm} and \ref{eq-ftj} applied to $F_{\rho}$ yield 
\begin{eqnarray}
\label{eq-ftFE} 
{\mathcal F}_{\rho}(T) &=& \frac{1}{2}\sum_{\bm q} \sum_{a,b} \sum_{u,v}
m_{\bm q}^{a,u} \left( nT\delta^{a,b}\delta^{u,v}
- {\mathcal J}^{ab}_{uv}({\bm q}) \right) m_{-{\bm q}}^{b,v} \nonumber \\
&-& \sum_{\bm q}\sum_{a,b}\sum_{u,v} \delta^{a,b}\delta^{u,v}
h_{\bm q}^{a,u} m_{-{\bm q}}^{b,v} 
- Tp\ln C_1 \ ,
\end{eqnarray}
where ${\mathcal F}_{\rho}(T)=F_{\rho}(T)/N_{\rm cell}$. A transformation to normal modes is
necessary to diagonalize ${\mathcal J}^{ab}_{uv}({\bm q})$. This is accomplished by
the use of Eq.~\ref{eq-nmodes}, or  
\begin{equation}
\label{eq-normmodes}
m_{\bm q}^{a,u} = \sum_{\alpha=1}^4 \sum_{\mu=1}^3 U^{a,\alpha}_{u,\mu}({\bm q}) 
\phi_{\bm q}^{\alpha,\mu} \ ,
\end{equation} 
where the Greek indices $(\alpha,\mu)$ label the normal modes. 
$U({\bm q})$ is the unitary matrix that diagonalizes 
${\mathcal J}({\bm q})$ in the sublattice space with eigenvalues $\lambda({\bm q})$,
\begin{equation} 
\label{eq-unitaryT}
U^{\dagger}({\bm q}){\mathcal J}({\bm q})U({\bm q}) = \lambda({\bm q}) \ ,  
\end{equation}
where in component form $U^{a,\alpha}_{u,\mu}({\bm q})$ represents the $(a,u)$ component of
the $(\alpha,\mu)$ eigenvector at ${\bm q}$ with eigenvalue $\lambda^{\alpha}_{\mu}({\bm q})$. 
The amplitudes of the normal modes are denoted by 
$\{\phi_{\bm q}\} = \{\sum_{\alpha,\mu} \phi_{\bm q}^{\alpha,\mu} \}$. 
Therefore, the MF free energy to quadratic order in the normal modes variables reads, 
\begin{eqnarray}
\label{eq-fenm}
{\mathcal F}_{\rho}(T) &=& \frac{1}{2} 
\sum_{\bm q}\sum_{\alpha,\mu} \phi_{\bm q}^{\alpha,\mu} \left( nT - 
\lambda^{\alpha}_{\mu}({\bm q}) \right) \phi_{-{\bm q}}^{\alpha,\mu} \\
&-& T \sum_{\bm q} \sum_{\alpha,\mu} {\tilde h}_{\bm q}^{\alpha,\mu} \phi_{-\bm q}^{\alpha,\mu} 
- Tp\ln C_1 \nonumber \ ,
\end{eqnarray}
where 
\[
{\tilde h}_{\bm q}^{\alpha,\mu} =  \frac{1}{T} \sum_{a,u} {\bm h}_{\bm q}^{a,u}
U^{a,\alpha}_{u,\mu}(-{\bm q}) \ .
\]
Note that for the Ising case, the indices representing the spin components $(u,v)$ 
and the corresponding modes $(\mu,\nu)$ are dropped from Eq.~\ref{eq-fenm}. 

The neutron-scattering scattering cross section for unpolarized neutrons in the 
dipole approximation is given by Eq.~\ref{eq-Xsect1} 
(Refs.~\onlinecite{jensen-book} and \onlinecite{lovesey-book}), or 
\begin{equation}
\frac{d\sigma(Q)}{d\Omega} = \frac{C [f(Q)]^{2}}{N_{\rm cell}} \sum_{i,j} \sum_{a,b}
\langle {\bm S}_{i\perp}^{a} \cdot {\bm S}_{j\perp}^{b} \rangle
\e^{\imath {\bm Q} \cdot {\bm R}_{ij}^{ab}}\ ,
\label{eq-Xsect-a} 
\end{equation}
where ${\bm Q}$ is the momentum transfer, i.e.,
${\bm Q} = {\bm G} + {\bm q}$, ${\bm G}$ is a 
fcc reciprocal lattice vector, ${\bm q}$ is 
a wave vector in the first Brillouin zone, and $C$ is
a constant.
The correlation function is between spin components perpendicular to the
vector ${\bm Q}$, 
\begin{eqnarray}
\label{eq-corr-perp-a}
\langle {\bm S}_{i\perp}^{a} \cdot {\bm S}_{j\perp}^{b} \rangle &=& 
\sum_{u,v} \left( {\hat n}_{\perp}^{u} \cdot {\hat n}_{\perp}^{v} \right) 
\langle S_{i}^{a,u} S_{j}^{b,v} \rangle  \\
&=& \sum_{u,v} \left ( \delta^{u,v} - \frac{Q^u Q^v}{|{\bm Q}|^2} \right )
\langle S_{i}^{a,u} S_{j}^{b,v} \rangle  \nonumber \ , 
\end{eqnarray}
where ${\hat n}_{\perp}^{u} = {\hat n}^{u} - ({\hat n}^{u} \cdot {\hat Q}){\hat Q}$ 
is strictly a geometric factor and $\hat{Q} = {\bm Q}/|{\bm Q}|$. For Ising spins 
one replaces ${\hat n}^{u}$ with ${\hat z}^{a}$ from Table 
\ref{tab-raza} and $S_i^{a,u}$ with $\sigma_i^a$. In order to proceed, 
the correlation function between spin variables, 
$\langle {\bm S}_{i\perp}^{a} \cdot {\bm S}_{j\perp}^{b} \rangle$, 
must be transformed to ${\bm q}$-space by use of Eq.~\ref{eq-ftm} and then 
to normal mode variables by application of Eq.~\ref{eq-normmodes}, one arrives at
\begin{eqnarray}  
\label{eq-correlnm-a}
\langle S_{i}^{a,u} S_{j}^{b,v} \rangle &=& 
\sum_{{\bm q},{\bm q}\prime}\sum_{\alpha,\beta}\sum_{\mu,\nu} 
\langle \phi_{\bm q}^{\alpha,\mu} \phi_{{\bm q}\prime}^{\beta,\nu} \rangle 
 \\ 
&\times & U^{\alpha, a}_{\mu, u}({\bm q}) U^{b,\beta}_{v,\nu}({\bm q}\prime)
\e^{-\imath {\bm q} \cdot {\bm R}_i^a - \imath {\bm q}\prime \cdot {\bm R}_j^b} 
\nonumber \ .
\end{eqnarray}
The correlation function 
$\langle \phi_{\bm q}^{\alpha,\mu} \phi_{{\bm q}\prime}^{\beta,\nu} \rangle$ 
can be calculated from a partition function defined in terms of the normal mode
amplitudes. The general definition reads, 
\begin{equation}
Z = {\rm Tr}\{\e^{-{\mathcal F}_{\rho}(T)/T} \} \ ,
\end{equation}
where ${\mathcal F}_{\rho}(T)$ is given by Eq.~\ref{eq-fenm}, and the trace is 
over all values of the normal mode amplitudes,
\[
{\rm Tr} \equiv \int_{-\infty}^{\infty} \prod_{{\bm q},\alpha,\mu} d\phi_{\bm q}^{\alpha,\mu} \ ,
\]
so one has,
\begin{equation}
Z = \prod_{{\bm q},\alpha,\mu} \int_{-\infty}^{\infty} d\phi_{\bm q}^{\alpha,\mu}
\e^{-\frac{1}{2}(n - \frac{\lambda^{\alpha}_{\mu}({\bm q})}{T}) |\phi_{{\bm q}}^{\alpha,\mu}|^2 
+ {\tilde h}_{\bm q}^{\alpha,\mu} \phi_{-\bm q}^{\alpha,\mu}} \ ,
\end{equation}
where a constant term has been dropped. The integral above is recast 
as a general Gaussian,
\begin{equation}
\int_{-\infty}^{\infty} d\phi \;  
\e^{-\frac{1}{2}A\phi^2 + B \phi} = \sqrt{{\frac{2\pi}{A}}} \e^{\frac{B^2}{2A}} \ ,
\end{equation}
where $A=(n - \frac{\lambda^{\alpha}_{\mu}({\bm q})}{T})$ and 
$B={\tilde h}_{\bm q}^{\alpha,\mu}$. 
Therefore, the final form of the partition function is,
\begin{eqnarray}
Z &=& \prod_{{\bm q},\alpha,\mu} Z^{\alpha,\mu}({\bm q}) \\ 
&=& \prod_{{\bm q},\alpha,\mu} \left[ \frac{2\pi}
{n - \frac{\lambda^{\alpha}_{\mu}({\bm q})}{T}}\right]^{(1/2)}
\e^{|{\tilde h}_{\bm q}^{\alpha,\mu}|^2/
2(n - \lambda^{\alpha}_{\mu}({\bm q})/T)} \; .
\nonumber 
\end{eqnarray}
The correlation function is now determined from derivatives of $Z$ with
respect to the fields ${\tilde h}_{\bm q}^{\alpha,\mu}$,
\begin{eqnarray}
\label{eq-Zcorrel}
\langle \phi_{\bm q}^{\alpha,\mu} \phi_{{\bm q}\prime}^{\beta,\nu} \rangle 
&=& \frac{1}{Z} 
\frac{\partial^2 Z}{\partial \tilde{h}_{\bm q}^{\alpha,\mu} \partial 
{\tilde h}_{{\bm q}\prime}^{\beta,\nu}}
\left |_{\tilde{h}_{\bm q}=0} \right . \\
&=& \frac{\delta_{{\bm q}+{\bm q}\prime,0}\; \delta^{\alpha,\beta}\; \delta^{\mu,\nu}}
{\left(n - \frac{\lambda_{\mu}^{\alpha}({\bm q})}{T}\right)} \nonumber \ .
\end{eqnarray}
Back substitution of the result from Eq.~\ref{eq-Zcorrel}, into Eqs.~\ref{eq-correlnm-a},
\ref{eq-corr-perp-a} and 
then into Eq.~\ref{eq-Xsect-a}, and finally imposing the properties of 
the Kronecker delta functions leaves
\begin{eqnarray}
\frac{1}{N_{\rm cell}} \frac{d\sigma(Q)}{d\Omega} &=& C [f(Q)]^{2} \sum_{\alpha,\mu} 
\sum_{a,b} \sum_{u,v} \frac{({\hat n}_{\perp}^{u} \cdot {\hat n}_{\perp}^{v})}
{\left(n - \lambda_{\mu}^{\alpha}({\bm q})/T\right)} \nonumber \\ 
&\times& U^{\alpha,a}_{\mu,u}({\bm q}) U^{b,\alpha}_{v,\mu}(-{\bm q}) 
\e^{\imath {\bm G} \cdot {\bm r}^{ab}} \ ,
\end{eqnarray}
where we have used the identity 
\[
\sum_{i,j} \e^{\imath ({\bm Q} - {\bm q}) \cdot {\bm R}_{ij}}
\e^{\imath ({\bm Q} - {\bm q}) \cdot {\bm r}^{ab}}
= N_{\rm cell}^2 \e^{\imath {\bm G} \cdot {\bm r}^{ab}} \ .
\]
We find it convenient to define the function,
\begin{eqnarray} 
\label{eq-scattfnH}
{\bm F}^{\alpha}_{\mu,\perp}({\bm q}) &=& \sum_{a,u} 
{\hat n}_{\perp}^{u}U^{\alpha,a}_{\mu,u}({\bm q}) 
\e^{\imath {\bm G} \cdot {\bm r}^a}   \\
&=& \sum_a \left \{{\bm U}^{\alpha,a}_{\mu}({\bm q}) 
- ({\bm U}^{\alpha,a}_{\mu}({\bm q}) \cdot {\hat Q}){\hat Q} 
\right \}\e^{\imath {\bm G} \cdot {\bm r}^a} \nonumber \ ,
\end{eqnarray}
where ${\bm U}^{\alpha,a}_{\mu}({\bm q}) 
= \sum_{u}{\hat n}^u U^{\alpha,a}_{\mu,u}({\bm q})$ 
(and therefore ${\bm F}^{\alpha}_{\mu,\perp}({\bm q})$) 
is a three component vector. The scattering cross section is written compactly as
\begin{equation}
\label{eq-nsXsectH-a}
\frac{1}{N_{\rm cell}} \frac{d\sigma(Q)}{d\Omega} = C [f(Q)]^{2} \sum_{\alpha,\mu} 
\frac{|{\bm F}^{\alpha}_{\mu,\perp}({\bm q})|^2}
{\left(n - \lambda_{\mu}^{\alpha}({\bm q})/T\right)} \ ,
\end{equation}
which is Eq.~\ref{eq-XsectH} with $n=3$. 
When considering \Is\ Ising spins, the arguments that follow 
Eq.~\ref{eq-correlnm-a} still 
hold, but the indices for the spin components ($u,v$) and corresponding 
normal modes ($\mu,\nu$) are dropped from the equations. The final expression for
the scattering cross section reads,
\begin{equation} 
\label{eq-nsXsectI-a}
\frac{1}{N_{\rm cell}} \frac{d\sigma(Q)}{d\Omega}  = C [f(Q)]^{2} \sum_{\alpha} 
\frac{|{\bm F}^{\alpha}_{\perp}({\bm q})|^2}
{\left(1 - \lambda^{\alpha}({\bm q})/T\right)} \ ,
\end{equation}
where ${\bm F}^{\alpha}_{\perp}({\bm q})$ is a three component vector given by,
\begin{equation}
\label{eq-scattfnI}
{\bm F}^{\alpha}_{\perp}({\bm q}) = \sum_{a} {\hat z}_{\perp}^{a}
U^{\alpha,a}({\bm q})  \e^{\imath {\bm G} \cdot {\bm r}^{a}} \ , 
\end{equation}
which is just Eq.~\ref{eq-XsectI}. We note that Eqs.~\ref{eq-nsXsectH-a} 
and \ref{eq-nsXsectI-a} are only valid for 
$T > \lambda_{\mu}^{\alpha}(\bm q_{\rm ord})/n$, where 
$\lambda_{\mu}^{\alpha}(\bm q_{\rm ord})
={\rm max}_{\bm q}\{ \lambda^{\rm max}({\bm q}) \}$ 
is the critical eigenvalue, a global maximum, which sets $T_c^{\rm MF}$ and 
defines the paramagnetic regime.
We also point out that the lattice structure and spin symmetry are 
contained in the respective ${\bm F}_{\perp}({\bm q})$-functions, 
Eqs.~\ref{eq-scattfnH} and \ref{eq-scattfnI}; 
these properties will be exploited in Appendix \ref{append:symm}
when we discuss symmetry allowed scattering patterns for Heisenberg and
Ising spins. 

%%%%%%%%%%%%%%%%
\subsection{Comment on the convention for the Fourier transform}
\label{append:nscatt-comment} 

Another convention for defining Fourier modes uses the
Bravais lattice points, ${\bm R}_i$, in the definitions \cite{FT-convention},     
\begin{equation}
\label{eq-ftmC2}
m_{i}^{a,u} = \sum_{\bm q} m_{\bm q}^{a,u} \e^{-\imath {\bm q} \cdot {\bm R}_{i}} \ ,
\end{equation}
and 
\begin{equation}
\label{eq-ftjC2}
{\mathcal J}^{ab}_{uv}(i,j) = \frac{1}{N_{\rm cell}}
\sum_{\bm q} {\mathcal J}^{a b}_{uv}({\bm q}) \e^{\imath {\bm q} \cdot {\bm R}_{ij}} \ .
\end{equation}
Hence, this convention differs from the one we employ by a simple phase factor,
$\exp{(\imath {\bm q} \cdot {\bm r}^{ab})}$ for ${\mathcal J}^{ab}_{uv}(i,j)$.  
The two approaches are equally valid and yield the same results; however, there
are a couple important differences in the above results when
this alternate convention is used. First, the interaction
matrix defined by the inverse of Eq.~\ref{eq-ftjC2} has complex entries. For nearest
neighbor interactions between sites $a=1$ and $b=2$, we have
\begin{equation}
{\mathcal J}^{a b}_{uv}({\bm q}) = {\mathcal J}^{1,2}_{uv} (1 + 
\e^{-\imath {\bm q} \cdot {\bm R}_{i,j}}) \ ,
\end{equation}
where the factor of $1$ arises because the sites $a=1$ and $b=2$ are
in the same tetrahedral basis unit, i.e., ${\bm R}_{i,j} = {\bm 0}$, 
but its symmetric equivalent has ${\bm R}_{i,j} \neq {\bm 0}$.
Next, the definition for the scattering cross section, Eq.~\ref{eq-Xsect-a},
holds, but the expression for the correlation function between normal mode
variables, Eq.~\ref{eq-correlnm-a}, contains the phase factor 
$\exp{(-\imath {\bm q} \cdot {\bm R}_{i} - \imath {\bm q}' \cdot {\bm R}_{j})}$.
Carrying the necessary steps through to an expression for scattering cross section, 
one finds that the factor $\exp{(\imath {\bm G} \cdot {\bm r}^a)}$ in 
Eqs.~\ref{eq-scattfnH} and \ref{eq-scattfnI} is replaced by 
$\exp{(\imath {\bm Q} \cdot {\bm r}^a)}$. Recall that momentum 
transfer and reciprocal lattice vector are related according to 
${\bm Q} = {\bm G} + {\bm q}$.
The presentation found in Refs.~\onlinecite{rbs-mft} and \onlinecite{reimers-mft}
(as noticed recently by Kadowaki et al.~\cite{hosno-kadowaki}) 
unintentionally mixes these two conventions. 

%%%%%%%%%%%%%%%%%%%%%%%%%%%%%%%%%%%%%%%%%%%%%%%%%%%%%%%%%%%%%%%%%%
\section{High temperature expansion of the Gaussian approximation \protect}
\label{append:htse}

We demonstrate that the equations for the neutron-scattering
cross section obtained from MFT at the Gaussian level, Appendix \ref{append:nscatt},
can also be formulated via a high temperature series expansion (HTSE) 
to lowest order in $\beta \equiv 1/T$, where $T$ is temperature in units of
$k_{\rm B}$. 
In contrast to VMFT, in a HTSE there is no appeal to any
simplifying approximation that changes the character of 
the density matrix 
and imposes constraints that keep the internal energy, ${\rm Tr}\{\rho H\}$, fixed. 
We follow our established convention of treating 
the general case of anisotropic Heisenberg spins 
while pointing out the specific differences for \Is\ Ising spins
when needed. 
The starting point is the Heisenberg Hamiltonian of Eq.~\ref{eq-Hheis}
\begin{equation}
H_H = - \frac{1}{2}\sum_{i,j} \sum_{a,b} \sum_{u,v}
{\mathcal J}^{ab}_{uv}(i,j) S_{i}^{a,u} S_{j}^{b,v}
\label{Aeq-Ham111}
\end{equation}
where ${\mathcal J}^{ab}_{uv}(i,j)$ contains both spin and coordinate
degrees of freedom and is defined by Eq.~\ref{eq-JH}. 

In the formula for the scattering cross section, Eq.\ref{eq-Xsect1}, 
one must calculate the perpendicular correlation function, 
\begin{equation}
\label{eq-corrperp-a}
\langle {\bm S}_{i\perp}^{a} \cdot {\bm S}_{j\perp}^{b} \rangle = 
\sum_{u,v} ({\hat n}_{\perp}^{u} \cdot {\hat n}_{\perp}^{v})
\langle S_i^{a,u} S_j^{b,v} \rangle \; ,
\end{equation}
which is just Eq. \ref{eq-corrperp}. 
We express the correlation function $\langle S_i^{a,u} S_j^{b,v} \rangle$
as a series expansion in cumulants \cite{htse-DG}, 
\begin{equation}
\label{eq-cumm}
\langle S_i^{a,u} S_j^{b,v} \rangle = \sum_{m=0}^{\infty}
\frac{(-\beta)^{m}}{m!} \langle S_i^{a,u} S_j^{b,v}
H^{m}_{H}\rangle_{c},
\end{equation}
where $\langle ... \rangle$ represents a thermal average with respect to
$H_H$ and $\langle ... \rangle_{c}$ represents the cumulant expansion of the
spin operators and $H_H$. Cumulants are evaluated as a trace over the
$T=0$ states, $\langle ... \rangle_{0}= {\rm Tr}\{ ... \}/{\rm Tr}\{1\}$, where
${\rm Tr}\{1\}={\tilde N}$ is a normalization factor, $(4\pi)^N$ for Heisenberg spins
and $(2)^N$ for Ising spins, $N$ is the total number of sites in the lattice. Therefore,
the correlation function to lowest order in $\beta$ is 
\begin{equation}
\label{eq-corr-cumm1}
\langle S_i^{a,u} S_j^{b,v} \rangle \approx
\langle S_i^{a,u} S_j^{b,v} \rangle_{c} -
\beta \langle S_i^{a,u} S_j^{b,v} H_H \rangle_{c} + \cdots
\end{equation}
For both Heisenberg and Ising Hamiltonians,  
any non-zero contribution to the correlation function must have
an even number of spin components per site (i.e., $(S_{i}^{a,u})^2$).  
The zeroth order term in $\beta$ is obtained trivially, 
\[
\langle S_i^{a,u} S_j^{b,v} \rangle_{c} = \langle S_i^{a,u} S_j^{b,v} \rangle_{o} 
= (1/n)\delta_{i,j}\delta^{a,b}\delta^{u,v} \; ,
\]
The first order contribution has two terms in the cumulant, 
\[
\langle S_i^{a,u} S_j^{b,v} H \rangle_{c} = \langle S_i^{a,u} S_j^{b,v} H \rangle_{o}
-  \langle S_i^{a,u} S_j^{b,v} \rangle_{o} \langle H \rangle_{o} \; ,
\]
but the second does not contribute because 
$\langle H \rangle_{o} \propto \langle S_l^{{\tilde a},{\tilde u}} 
S_m^{{\tilde b},{\tilde v}} \rangle_{o} = 0$, spins in the Hamiltonian
can not be at the same site. The first term yields the result
\begin{eqnarray}
\langle S_i^{a,u} S_j^{b,v} H \rangle_{o} &=& -\frac{1}{2}
\sum_{l,m}\sum_{{\tilde a},{\tilde b}}\sum_{{\tilde u},{\tilde v}}
{\mathcal J}^{{\tilde a}{\tilde b}}_{{\tilde u}{\tilde v}}(i,j) 
\langle S_i^{a,u} S_j^{b,v} S_l^{{\tilde a},{\tilde u}} S_m^{{\tilde b},{\tilde v}} \rangle
\nonumber \\
&=& -(1/n^2){\mathcal J}^{ab}_{uv}(i,j) \; .
\end{eqnarray}
Therefore, Eq.~\ref{eq-corr-cumm1} can be rewritten as 
\begin{equation}
\label{eq-corr-cumm2}
\langle S_i^{a,u} S_j^{b,v} \rangle \approx
(1/n) \delta_{i,j}\delta^{a,b}\delta^{u,v} +
(\beta/n^2){\mathcal J}^{ab}_{uv}(i,j)
\end{equation}

We obtain an expression for the scattering cross section by substituting
the result of Eq.~\ref{eq-corr-cumm2} into
Eq.~\ref{eq-corrperp-a}, and then that result into Eq.~\ref{eq-Xsect1}; we 
use the identity ${\bm Q} = {\bm q} + {\bm G}$ 
and the definition of the Fourier transform of
${\mathcal J}^{ab}_{uv}(i,j)$, Eq.~\ref{eq-FTJ}, to arrive at, 
\begin{eqnarray}
\label{eq-XsectHTS}
\frac{1}{N_{\rm cell}}\frac{d\sigma(Q)}{d\Omega}& = & (C/n)[f(Q)]^{2}
\sum_{a,b}\sum_{u,v}
({\hat n}_{\perp}^{u} \cdot {\hat n}_{\perp}^{v}) \\
&\times& \left (\delta^{a,b}\delta^{u,v} + \frac{\beta}{n} 
{\mathcal J}^{ab}_{uv}(-{\bm q}) \right ) 
\e^{\imath {\bm G} \cdot {\bm r}^{ab}} \nonumber \; ,
\end{eqnarray}
where $C$ is a constant. Note, a factor of $1/N_{\rm cell}$ has been absorbed into
the expression of the Fourier transform of the correlation function, 
Eq.~\ref{eq-corr-cumm2}, in order to remain consistent with our use of 
Eq.~\ref{eq-ftFE} in the derivation of the mean-field neutron-scattering 
cross section, Eq.~\ref{eq-nsXsectH-a}.
The interaction matrix, ${\mathcal J}^{ab}_{uv}(-{\bm q})$, 
in Eq.~\ref{eq-XsectHTS} is not diagonal. In order to calculate the differential 
cross section for the pyrochlore lattice, or any lattice with a basis, 
we must diagonalize ${\mathcal J}^{ab}_{uv}(-{\bm q})$.
This is done with unitary matrix $U({\bm q})$, or $U^{a,\alpha}_{u,\mu}({\bm q})$ 
in component form. The first term on the right hand side of 
Eq.~\ref{eq-XsectHTS} follows 
directly from the definition of $U({\bm q})$,
\begin{equation}
\label{eq-part1}
\sum_{\alpha,\mu} U^{a,\alpha}_{u,\mu}(-{\bm q}) U^{\alpha,b}_{\mu,v}({\bm q}) 
= {\bm I} = \delta^{a,b}\delta^{u,v} \ ,
\end{equation}
where ${\bm I}$ is the appropriate identity matrix.
The transformation of the second term uses Eq.~\ref{eq-unitaryT}, 
and solves for ${\mathcal J}({\bm q})$,
\begin{equation}
{\mathcal J}({\bm q}) = U({\bm q})\lambda({\bm q})U^{\dagger}({\bm q}) \nonumber \ ,
\end{equation}
or in component form,  
\begin{equation}
\label{eq-part2}
{\mathcal J}^{ab}_{uv}(-{\bm q}) = \sum_{\alpha,\mu} \lambda^{\alpha}_{\mu}({\bm q})
U^{a,\alpha}_{u,\mu}(-{\bm q}) U^{\alpha,b}_{\mu,v}({\bm q}) \ ,
\end{equation}
where $\lambda^{\alpha}_{\mu}(-{\bm q})=\lambda^{\alpha}_{\mu}({\bm q})$.
Using the results from Eqs.~\ref{eq-part1} and \ref{eq-part2}, the 
expression for the scattering cross section becomes,
\begin{eqnarray}
& &\frac{1}{N_{\rm cell}}\frac{d\sigma(Q)}{d\Omega}=(C/n)[f(Q)]^{2}
\sum_{\alpha,\mu}\sum_{a,b} \sum_{u,v}
(\hat{n}_{\perp}^{u} \cdot \hat{n}_{\perp}^{v}) \nonumber \\
& &\times \left(1 + \frac{\beta}{n} \lambda^{\alpha}_{\mu}({\bm q}) \right) 
U^{a,\alpha}_{u,\mu}(-{\bm q}) U^{\alpha,b}_{\mu,v}({\bm q})
\e^{\imath {\bm G} \cdot {\bm r}^{ab}}
\end{eqnarray}
In the high temperature limit ($\beta \rightarrow 0$), one can
write 
\[
\left(1 + \frac{\beta}{n}\lambda^{\alpha}_{\mu}({\bm q}) \right) \approx
\left(1 - \frac{\beta}{n}\lambda^{\alpha}_{\mu}({\bm q}) \right)^{-1} \ ,
\]
and the mean-field result is recovered,
\begin{equation}
\label{eq-Xs1}
\frac{1}{N_{\rm cell}}\frac{d\sigma(Q)}{d\Omega} 
= C [f(Q)]^{2}\sum_{\alpha,\mu}
\frac{\left|{\bm F}^{\alpha}_{\mu,\perp}({\bm q}) \right|^2}
{\left(n - \beta \lambda^{\alpha}_{\mu}({\bm q}) \right)} \ ,
\end{equation}
where ${\bm F}^{\alpha}_{\mu,\perp}({\bm q})$ is given by
Eq.~\ref{eq-scattfnH} with $U^{a,\alpha}_{u,\mu}(-{\bm q})=U^{\alpha,a}_{\mu,u}({\bm q})$. 
For a \Is\ Ising model, the indices 
$(u,v)$ and $(\mu,\nu)$ are dropped from our presentation, $n=1$, and 
${\bm F}^{\alpha}_{\perp}({\bm q})$ is given by Eq.~\ref{eq-scattfnI}. 

%%%%%%%%%%%%%%%%%%%%%%%%%%%%%%%%%%%%%%%%%%%%%
\section{Ewald Equations \protect}
\label{append:ewald}

Here we treat the dipole-dipole term in ${\mathcal J}({\bm q})$ 
via the Ewald method \cite{ewald}. In MFT one works in the thermodynamic
limit, so one has an infinite lattice sum. Within the Ewald approach,
one recasts this infinite and conditionally convergent series as two 
finite absolutely convergent (rapidly converging) sums 
\cite{born-huang,deleeuw}. The application of Ewald's ideas to 
the Fourier transformed dipole-dipole interaction is equivalent 
to the method of long wave lengths presented in 
Ref.~\onlinecite{born-huang}.  

The general expression of the Fourier transformed dipole-dipole lattice sum is 
\begin{equation}   
{\mathcal A}({\bm q}) = \sum_{i}{\textstyle ^{'}} \sum_{\stackrel{a,b}{u,v}}
{\mathcal A}^{ab}_{uv}\, \e^{-\imath {\bm q} \cdot {\bm R}_{ij}^{ab}} \ ,
\label{eq-Hdip}
\end{equation}
where 
\begin{equation}   
{\mathcal A}^{ab}_{uv} = \frac{{\hat n}^{u} \cdot 
{\hat n}^{v}}{|{\bm R}_{ij}^{ab}|^3} - 
\frac{3 ({\hat n}^{u}\cdot {\bm R}_{ij}^{ab}) 
({\hat n}^{v}\cdot {\bm R}_{ij}^{ab})}{|{\bm R}_{ij}^{ab}|^5} \ .
\label{eq-Aabuv}
\end{equation}
The conventions for indices and vectors are described in Section \ref{sec:mft}, 
and ${\mathcal A}({\bm q})$ is a $12\times12$ matrix.
The sum $\sum_{i}^{'}$ is over all ${\bm R}_{ij}^{ab}$ except the terms  
${\bm R}_{ij}^{ab}={\bm 0}$. To implement the Ewald method, 
we rewrite Eq.~\ref{eq-Aabuv} in the following form \cite{AF},
\begin{equation}
{\mathcal A}^{ab}_{uv} = 
-(\hat{n}^{u}\cdot \nabla_{x})(\hat{n}^{v}\cdot \nabla_{x}) 
\left \{\frac{1}{|{\bm R}_{ij}^{ab} - {\bm x}|} 
\right \}_{{\bm x}={\bm 0}} \ .
\label{eq-Aabuv2}
\end{equation}
Equation \ref{eq-Aabuv2} is used in Eq.~\ref{eq-Hdip}, and in terms of 
components one has,  
\begin{equation}
{\mathcal A}^{ab}_{uv}({\bm q}) = 
-(\hat{n}^{u}\cdot \nabla_{x})(\hat{n}^{v}\cdot \nabla_{x}) 
\left \{\sum_{i}{\textstyle ^{'}} \frac{\e^{-\imath {\bm q} \cdot {\bm R}_{ij}^{ab}}}
{|{\bm R}_{ij}^{ab} - {\bm x}|} \right \}_{{\bm x}={\bm 0}} \ .
\label{eq-Aq1}
\end{equation}

The goal of the Ewald method is to rewrite Eq.~\ref{eq-Aq1}, a conditionally convergent
series, as two absolutely convergent series, one in real space and the other 
in reciprocal space ($k$-space). We begin by writing the sum inside the brackets as 
a sum over all ${\bm R}_{ij}^{ab}$, the result is, 
\begin{eqnarray}
{\mathcal A}^{ab}_{uv}({\bm q}) &=& 
-(\hat{n}^{u}\cdot \nabla_{x})(\hat{n}^{v}\cdot \nabla_{x}) 
\left \{\sum_{i} \frac{\e^{-\imath {\bm q} \cdot {\bm R}_{ij}^{ab}}}
{|{\bm R}_{ij}^{ab} - {\bm x}|} \right \}_{{\bm x}={\bm 0}} \nonumber \\
&+& \delta^{ab} (\hat{n}^{u}\cdot \nabla_{x})(\hat{n}^{v}\cdot \nabla_{x})  
\left \{\frac{1}{|{\bm x}|}\right \}_{{\bm x}={\bm 0}} \; .
\label{eq-Aq1b}
\end{eqnarray}
Next, the definition of a Gaussian integral (also 
a gamma function identity \cite{arfken}), 
\[
\frac{1}{|{\bm R}|}=\frac{2}{\sqrt{\pi}}\int_{0}^{\infty} \e^{-t^2 R^2} dt \ ,
\]
is used rewrite the point source term, 
$1/|{\bm R}_{ij}^{ab} - {\bm x}|$, in Eq.~\ref{eq-Aq1b}. 
The Fourier transformed dipole-dipole lattice sum now reads,
\begin{eqnarray}
{\mathcal A}^{ab}_{uv}({\bm q}) &=& 
-(\hat{n}^{u}\cdot \nabla_{x})(\hat{n}^{v}\cdot \nabla_{x})
\int_{0}^{\infty} dt\ \frac{2}{\sqrt{\pi}} \e^{-\imath {\bm q} \cdot {\bm x}} \nonumber \\
&\times&\left \{ 
\sum_{i} \e^{-t^2|{\bm R}_{ij}^{ab} - {\bm x}|^2 - \imath{\bm q} \cdot ({\bm R}_{ij}^{ab}- {\bm x})}
\right \}_{{\bm x}={\bm 0}}
\nonumber \\
&+&\delta^{ab} (\hat{n}^{u}\cdot \nabla_{x})(\hat{n}^{v}\cdot \nabla_{x})  
\left \{\frac{1}{|{\bm x}|}\right \}_{{\bm x}={\bm 0}} \; .
\label{eq-Aq2}
\end{eqnarray}
The integral in Eq.~\ref{eq-Aq2} is divided into two regions, 
$[0,\alpha]$ and $[\alpha,\infty)$. It is from this decomposition that the 
real space ($[\alpha,\infty)$) and $k$-space ($[0,\alpha]$) series will arise.
We note that the series resulting from the $[\alpha,\infty)$ integral will 
have a divergence at ${\bm R}_{ij}^{ab}={\bm 0}$, hence this term is treated separately. 
The range of integration is controlled by $\alpha$; it has units of inverse distance
and will play the role of a convergence parameter in the final series. Equation  
\ref{eq-Aq2} now reads, 
\begin{equation}
{\mathcal A}^{ab}_{uv}({\bm q}) = W^{ab}_{uv}({\bm q}) + X^{ab}_{uv}({\bm q}) + 
Y^{ab}_{uv} \ , 
\label{eq-Aq3}
\end{equation}
where 
\begin{eqnarray}
\label{eq-W}
W^{ab}_{uv}({\bm q}) &=& -(\hat{n}^{u}\cdot \nabla_{x})(\hat{n}^{v}\cdot \nabla_{x})
\int_{0}^{\alpha} dt\ \frac{2}{\sqrt{\pi}} \e^{-\imath {\bm q} \cdot {\bm x}} \nonumber \\
&\times&\left \{ \sum_{i} \e^{-t^2|{\bm R}_{ij}^{ab} - {\bm x}|^2 - \imath{\bm q} \cdot 
({\bm R}_{ij}^{ab}- {\bm x})} \right \}_{{\bm x}={\bm 0}} \ , \\
\label{eq-X}
X^{ab}_{uv}({\bm q}) &=& -(\hat{n}^{u}\cdot \nabla_{x})(\hat{n}^{v}\cdot \nabla_{x})
\int_{\alpha}^{\infty} dt\ \frac{2}{\sqrt{\pi}} \e^{-\imath {\bm q} \cdot {\bm x}} \nonumber \\
&\times&\left \{ \sum_{i}{\textstyle ^{'}} 
\e^{-t^2|{\bm R}_{ij}^{ab} - {\bm x}|^2 - \imath{\bm q} \cdot 
({\bm R}_{ij}^{ab}- {\bm x})} \right \}_{{\bm x}={\bm 0}} \ , \\
\label{eq-Y}
Y^{ab}_{uv} &=& \delta^{ab}(\hat{n}^{u}\cdot \nabla_{x})(\hat{n}^{v}\cdot \nabla_{x})
\nonumber \\
& &\left \{\frac{1}{|{\bm x}|} - \frac{2}{\sqrt{\pi}} \int_{\alpha}^{\infty} 
\e^{-t^2|{\bm x}|^2} dt \right \}_{{\bm x}={\bm 0}} \ .
\end{eqnarray}
We treat the expressions for $W^{ab}_{uv}({\bm q})$, $X^{ab}_{uv}({\bm q})$, 
and $Y^{ab}_{uv}$ in succession.

For $W^{ab}_{uv}({\bm q})$, the sum inside the brackets is a periodic function in 
${\bm x}$. Therefore, it can be expressed as a Fourier series,
\begin{equation}
f(x) = \sum_{i} \e^{-t^2|{\bm R}_{ij}^{ab} - {\bm x}|^2 - \imath{\bm q} \cdot 
({\bm R}_{ij}^{ab}- {\bm x})}
= \sum_{\bm k} g_{\bm k} \e^{\imath {\bm k} \cdot {\bm x}} .
\label{eq-fs}
\end{equation}
Solving for $g_{\bm k}$ one has
\begin{equation}
g_{{\bm k}={\bm G}} = \frac{4\pi}{v}
\frac{\e^{-\imath {\bm G} \cdot {\bm r}^{ab}}}{|{\bm q}-{\bm G}|^3} F(z) ,
\label{eq-gk}
\end{equation}
where ${\bm G}$ is a reciprocal lattice vector, $v$ is the volume of the unit cell, 
\begin{equation}
\label{eq-Fz}
F(z) = \int_{0}^{\infty} y \ \sin(y) \ \e^{-z^2 y^2}\ dy  = \frac{\sqrt{\pi}}{4z^3} \e^{-1/4z^2} 
\end{equation}
and $z=t/|{\bm q}-{\bm G}|\;$ (Ref.~\onlinecite{comment-integral}). 
One now has the following identity for $f(x)$, 
\begin{eqnarray}
f(x) &=& \sum_{i} \e^{-t^2|{\bm R}_{ij}^{ab} - {\bm x}|^2 - \imath{\bm q} \cdot 
({\bm R}_{ij}^{ab}- {\bm x})}
\nonumber \\
&=& \frac{4\pi}{v} \sum_{\bm G}
\frac{\e^{-\imath {\bm G} \cdot ({\bm r}^{ab}-{\bm x})}}{|{\bm q}-{\bm G}|^3} F(z) \ .
\label{eq-fs2}
\end{eqnarray}
Substituting Eq.~\ref{eq-fs2} into Eq.~\ref{eq-W}, differentiating, and imposing the
limit on ${\bm x}$ yields, 
\begin{eqnarray}
\label{eq-W2}
W^{ab}_{uv}({\bm q}) &=& \frac{4\pi}{v} \sum_{\bm G} 
\frac{[{\hat n}^{u}\cdot ({\bm q}-{\bm G})]
[{\hat n}^{v}\cdot ({\bm q}-{\bm G})]}{|{\bm q}-{\bm G}|^3}
\e^{-\imath {\bm G} \cdot {\bm r}^{ab}} \nonumber \\
&\times& \frac{2}{\sqrt{\pi}} \int_{0}^{\alpha} dt\ 
F(t/|{\bm q}-{\bm G}|) \ .
\end{eqnarray}
The integral over $[0,\alpha]$ is readily performed by using the result from 
Eq.~\ref{eq-Fz}. Therefore, the reciprocal space sum in the Ewald decomposition 
reads, 
\begin{eqnarray}
\label{eq-W3}
W^{ab}_{uv}({\bm q}) &=& \frac{4\pi}{v} \sum_{\bm G} 
\frac{[{\hat n}^{u}\cdot ({\bm q}-{\bm G})]
[{\hat n}^{v}\cdot ({\bm q}-{\bm G})]}{|{\bm q}-{\bm G}|^2} \nonumber \\
&\times& \e^{-|{\bm q}-{\bm G}|^2/4\alpha^2} \e^{-\imath {\bm G} \cdot {\bm r}^{ab}}
 \ ,
\end{eqnarray}
where the sum is over all reciprocal lattice vectors ${\bm G}$. We note, however, 
the series for $W^{ab}_{uv}({\bm q})$ has a nonanalytic term at ${\bm G}={\bm 0}$ 
when at the zone center, ${\bm q}={\bm 0}$. This point is discussed below. 

The expression for $X^{ab}_{uv}({\bm q})$, Eq.~\ref{eq-X}, can be rearranged 
to obtain an identifiable integral. By reversing the sum and integral in
Eq.~\ref{eq-X} we obtain, 
\begin{eqnarray}
\label{eq-X2}
X^{ab}_{uv}({\bm q}) &=& -(\hat{n}^{u}\cdot \nabla_{x})(\hat{n}^{v}\cdot \nabla_{x}) 
\sum_{i}{\textstyle ^{'}} \e^{-\imath{\bm q} \cdot {\bm R}_{ij}^{ab}} \nonumber \\ 
&\times&\frac{2}{\sqrt{\pi}} \int_{\alpha}^{\infty} dt\ 
\e^{-t^2|{\bm R}_{ij}^{ab} - {\bm x}|^2} \left. \right |_{{\bm x}={\bm 0}}
 \ .
\end{eqnarray} 
The integral in Eq.~\ref{eq-X2} can be expressed as  
a complementary error function \cite{arfken},
\[
{\rm erfc}(z) = \frac{2}{\sqrt{\pi}} \int_{z}^{\infty} \e^{-x^2} dx \ .
\] 
A final form for $X^{ab}_{uv}({\bm q})$ is obtained by first applying the
differential operators in Eq.~\ref{eq-X2}, followed by taking 
the limit ${\bm x} \rightarrow 0$, and then integrating to get, 
\begin{equation}
\label{eq-X3}
X^{ab}_{uv}({\bm q}) = \sum_{i}{\textstyle ^{'}} 
\left ( S1^{ab}_{uv}({\bm R}_{ij}^{ab}) - 
S2^{ab}_{uv}({\bm R}_{ij}^{ab}) \right )
\e^{-\imath{\bm q} \cdot {\bm R}_{ij}^{ab}} \ , 
\end{equation} 
where 
\begin{eqnarray}
\label{eq-S1}
S1^{ab}_{uv}({\bm R}_{ij}^{ab}) &=& ({\hat n}^{u}\cdot {\hat n}^{v})
\\
&\times& \left \{ \frac{2\alpha}{\sqrt{\pi}}
\frac{\e^{-\alpha^2 |{\bm R}_{ij}^{ab}|^2}}{|{\bm R}_{ij}^{ab}|^2} + 
\frac{{\rm erfc}(\alpha |{\bm R}_{ij}^{ab}|)}{|{\bm R}_{ij}^{ab}|^3} \right \} \; , \nonumber
\end{eqnarray}
\begin{eqnarray}
\label{eq-S2}
S2^{ab}_{uv}({\bm R}_{ij}^{ab}) &=&
({\hat n}^{u}\cdot {\bm R}_{ij}^{ab})({\hat n}^{v}\cdot {\bm R}_{ij}^{ab}) \\
&\times& \left \{ \left [ \frac{4\alpha^3}{\sqrt{\pi} |{\bm R}_{ij}^{ab}|^2} 
+ \frac{6\alpha}{\sqrt{\pi} |{\bm R}_{ij}^{ab}|^4}\right ] \e^{-\alpha^2 |{\bm R}_{ij}^{ab}|^2}
\right . \nonumber \\
&+& \left . \frac{3\ {\rm erfc}(\alpha |{\bm R}_{ij}^{ab}|)}{|{\bm R}_{ij}^{ab}|^5} \right \}
\; . \nonumber
\end{eqnarray}
Equations \ref{eq-X3}-\ref{eq-S2} form the real space sum in the Ewald decomposition
of the dipole-dipole interaction. Note that the sum in Eq.~\ref{eq-X3} is over all 
Bravais lattice displacement vectors ${\bm R}_{ij}$, with $j$ fixed, except 
${\bm R}_{ij}={\bm 0}$. Hence, the real space Ewald series is analytic everywhere. 

In treating the singular terms in Eq.~\ref{eq-Y}, one 
applies differential operators first to get,
\begin{eqnarray}
\label{eq-Y2}
Y^{ab}_{uv} &=& \lim_{{\bm x}\rightarrow {\bm 0}} \delta^{ab} \\ 
&\times& \left \{ - \frac{({\hat n}^{u}\cdot {\hat n}^{v})}{|{\bm x}|^3} 
+ \frac{3 ({\hat n}^{u}\cdot {\bm x})({\hat n}^{v} \cdot {\bm x})} 
{|{\bm x}|^5} \right . \nonumber \\
&+& \left . \left ( S1^{ab}_{uv}({\bm x}) - S2^{ab}_{uv}({\bm x}) \right )
\right \} \nonumber \ ,
\end{eqnarray}
where $S1^{ab}_{uv}({\bm x})$ and $S2^{ab}_{uv}({\bm x})$ are given by 
Eqs.~\ref{eq-S1} and \ref{eq-S2}, respectively, with ${\bm R}_{ij}^{ab}$ 
replaced by ${\bm x}$. To evaluate the limit in Eq.~\ref{eq-Y2}, one 
expands the exponential function to $O({\bm x}^2)$ and the complementary
error function to order $O({\bm x}^3)$. The result is the constant, 
\begin{equation}
\label{eq-Y3}
Y^{ab}_{uv} = - \frac{4\alpha^3}{3\sqrt{\pi}} 
({\hat n}^{u} \cdot {\hat n}^{v}) \delta^{a,b} \ . 
\end{equation}
Collecting the results of Eqs.~\ref{eq-W3}, \ref{eq-X3}-\ref{eq-S2} 
and \ref{eq-Y3}, we write the Ewald representation of the ${\bm q}$-dependent 
dipole-dipole interaction as,
\begin{eqnarray}
\label{eq-Aq4}
{\mathcal A}^{ab}_{uv}({\bm q}) &=& -\frac{4\alpha^3}{3\sqrt{\pi}}
(\hat{n}^{u}\cdot \hat{n}^{v}) \delta^{a,b}  \\
&+& \frac{4\pi}{v}\sum_{\bm G} K_{uv}({\bm q}-{\bm G}) 
\e^{-|{\bm q}-{\bm G}|^2/4\alpha^2} \e^{-\imath {\bm G} \cdot {\bm r}^{ab}} \nonumber
\\
&+& \sum_{i}{\textstyle ^{'}} \left ( 
S1^{ab}_{uv}({\bm R}_{ij}^{ab}) - 
S2^{ab}_{uv}({\bm R}_{ij}^{ab}) \right ) 
\e^{-\imath{\bm q} \cdot {\bm R}_{ij}^{ab}} \; , \nonumber 
\end{eqnarray}
where 
\begin{equation}
K_{uv}({\bm q}-{\bm G}) = 
\frac{[{\hat n}^{u}\cdot ({\bm q}-{\bm G})]
[{\hat n}^{v}\cdot ({\bm q}-{\bm G})]}{|{\bm q}-{\bm G}|^2} \; 
\end{equation}

In our derivation of the Ewald equations there is no reference to a specific 
lattice structure. Therefore, 
the Ewald results encapsulated in Eq.~\ref{eq-Aq4} hold for any lattice 
described by a set of translation vectors $\{{\bm R}_{ij}^{ab}\}$. 
Through the unit vectors ${\hat n}^{u}$ (where local quantization axes 
can be treated by including a sublattice index, i.e., ${\hat n}^{a,u}$), 
the freedom to define the spin symmetry (e.g., Heisenberg, XY, Ising) 
has been ensured, too. For the work discussed in this article, we 
consider both Heisenberg and \Is\ Ising spins on the 
pyrochlore lattice. For Heisenberg spins,  
${\mathcal A}^{ab}_{uv}({\bm q})$ is calculated for all sublattices
$(a,b)$ and spin components $(u,v)$, the resulting 
${\mathcal A}({\bm q})$ is a $12\times 12$ symmetric matrix 
contribution to ${\mathcal J}({\bm q})$. For \Is\ Ising spins, the sums over
spin components are dropped and the local quantization vectors are substituted, ${\hat z}^a$.
One calculates ${\mathcal A}^{ab}({\bm q})$ for all sublattices $(a,b)$,  
resulting in ${\mathcal A}({\bm q})$ a symmetric $4\times 4$ contribution to 
${\mathcal J}({\bm q})$. For each pyrochlore model, ${\mathcal A}({\bm q})$ is determined at
every ${\bm q}$-point in a mesh that covers the first Brillouin zone in the
$(hhl)$ plane. These matrices are stored and then used in the formation 
of ${\mathcal J}({\bm q})$ to calculate the
neutron-scattering cross section, Eq.~\ref{eq-XsectH} or Eq.~\ref{eq-XsectI}, 
for a specified set of interaction parameters (i.e., $J$, $D$, $\Delta$, $T$).   
Because ${\mathcal A}({\bm q})$ is calculated only for ${\bm q}$ in the first
zone, the term ${\mathcal K}_{uv}({\bm q}-{\bm G})$ in Eq.~\ref{eq-Aq4} 
is ill defined at ${\bm q}={\bm G}=0$. We discuss the small ${\bm q}$  
behavior of the Ewald equations below. 

The parameter $\alpha$ used to divide the integral in Eq.~\ref{eq-Aq2} functions
as a convergence parameter in the Ewald sums, Eq.~\ref{eq-Aq4}. Although, the
result of ${\mathcal A}^{ab}_{uv}({\bm q})$ is independent of the
value of $\alpha$, in practice one chooses an $\alpha$ so that both real and
reciprocal sums converge rapidly. Note that the convergence of the 
real space sum, Eqs.~\ref{eq-X3}-\ref{eq-S2}, 
is enhanced by a large value for $\alpha$, while 
the convergence of the reciprocal sum, Eq.~\ref{eq-W3}, 
is improved for a small $\alpha$. In choosing a convergence parameter, we 
followed Ref.~\onlinecite{CK} and set $\alpha=\sqrt{\pi/v}$, where $v$ 
is the volume of the unit cell. For a pyrochlore lattice defined in the
rhombohedral basis with a cubic cell size of ${\bar a}$, we used $v={\bar a}^3/4$.    
The real and reciprocal space sums converged at about the same rate for this value
of $\alpha$. We obtained similar results for ${\mathcal A}^{ab}_{uv}({\bm q})$
using $\alpha=\sqrt{\pi/2v}$ and $\alpha=\sqrt{2\pi/v}$. 
Our Ewald results were checked by comparing the maximum eigenvalues 
of ${\mathcal A}({\bm q})$ to those generated from a direct lattice sum 
of ${\mathcal A}({\bm q})$ out to some cutoff distance $r_c$. Comparisons 
were done for the bcc and fcc lattices. 
We also performed tests of our Ewald equations for the pyrochlore lattice
by calculating the soft-mode spectrum of ${\mathcal A}({\bm q})$ in the 
spin-ice regime, e.g., $D=1$. Ewald results along 
$(00l)$ in the first Brillouin zone 
were compared to calculations
with the dipolar sum cutoff at different maximum separation distances $r_c$.
The cutoff results approach the Ewald results as $r_c$ increases. 
This spectrum of eigenvalues agrees well with the spectrum generated
from a direct lattice sum for ${\mathcal A}({\bm q})$ 
with a cutoff distance of $r_c=1000$, Fig.~$6$ in Ref.~\onlinecite{SI-mft}. 
The Ewald method eliminates the ripples in the soft-mode 
spectrum of ${\mathcal A}({\bm q})$ by effectively taking the range of 
interaction to infinity. 

The reciprocal space sum in Eq.~\ref{eq-Aq4} has a nonanalytic term 
at the point ${\bm q}={\bm 0}$ in the first Brillouin zone. 
If we consider the ${\bm G}={\bm 0}$ contribution to 
Eq.~\ref{eq-W3}, we have,
\begin{equation}
\label{eq-WG0_1}
W^{ab}_{uv}({\bm q},{\bm G}={\bm 0}) = \frac{4\pi}{v} \frac{({\hat n}^{u} \cdot {\bm q})
({\hat n}^{v} \cdot {\bm q})}{|{\bm q}|^2}\ \e^{-|{\bm q}|^2/4\alpha^2} 
\ .
\end{equation}
In the limit of small ${\bm q}$ the exponential is expanded to yield,
\begin{equation}
\label{eq-WG0_2}
W^{ab}_{uv}({\bm q},{\bm G}={\bm 0}) \approx \frac{4\pi}{v} \frac{({\hat n}^{u} \cdot {\bm q})
({\hat n}^{v} \cdot {\bm q})}{|{\bm q}|^2}\ 
\left (1 - \frac{|{\bm q}|^2}{4\alpha^2} \right )  
\ ,
\end{equation}
where in the limit ${\bm q} \rightarrow {\bm 0}$ the value of 
\[ 
\frac{4\pi}{v} \frac{({\hat n}^{u} \cdot {\bm q})
({\hat n}^{v} \cdot {\bm q})}{|{\bm q}|^2}
\]
depends on the direction in which one approaches the zone center. 
The nonanalytic term can be related to the macroscopic field
of the dipoles and is shape dependent, see Section $30$ 
in Ref.~\onlinecite{born-huang}. We drop this term from our
calculation to obtain a completely smooth spectrum all the
way to ${\bm q}=0,0,0$. The physics of spin-ice is not affected by this
omission because all modes contribute to the PM scattering 
with ${\bm q}=0,0,1$ going critical at $T_{c}^{\rm MF}$. 
The case of \Tb\ is more subtle because it is the 
${\bm q}=0,0,0$ soft mode that goes critical. However, our 
focus here is not the ordered state of \Tb , where a ${\bm q}={\bm 0}$ 
ordered state is expected for a pyrochlore AFM with either 
\Is\ Ising \cite{111-moessner-prb,tbtio-gingras1,dipSImodel1} 
or Heisenberg \cite{gdtio-palmerchalker} spins. Instead, we are concerned with
understanding the physics in the paramagnetic regime of this 
system as a first step toward unraveling the mystery surrounding 
the failure of \Tb\ to order at $50\; {\rm mK}$.  

%%%%%%%%%%%%%%%%%%%%%%%%%%%%%%%%%%%%%%%%%%%%%%%%%%%%%%%%%%%%%%%
\section{Symmetry excluded scattering \protect}
\label{append:symm}

The paramagnetic neutron-scattering spectrum of \Tb\ in the $(hhl)$ plane 
contains a strong but broad region of intensity about ${\bm Q}=0,0,2$ 
with no discernible 
correlations near the zone center, ${\bm Q}=0,0,0$ (Ref.~\onlinecite{tbtio-gardner2}). 
In this appendix, we put forward arguments based only on the structure of 
the lattice and the symmetry of spin space to demonstrate that the PM scattering 
intensity profile described above can not be realized by 
\Is\ Ising spins on the pyrochlore lattice but is allowed
if the spins are Heisenberg like.

For the Ising pyrochlores, the map of scattering intensity is determined 
by the function ${\bm F}_{\perp}^{\alpha}({\bm q})$, Eq.~\ref{eq-FperpI},
which contains only information on the symmetry of the lattice through 
the eigenvectors, $U^{a,\alpha}({\bm q})$, and the phase factor, 
$\exp{(\imath {\bm G} \cdot {\bm r}^a)}$, 
and the symmetry of spin space
through the local quantization axis, ${\bm z}^a$. 
We consider a unit tetrahedron with scattering vectors ${\bm Q}$ restricted 
to the $(00l)$ direction. To handle the situation near the origin, we express all 
${\bm Q}$ as a small displacement from a reciprocal lattice vector, i.e., 
${\bm Q}={\bm G}+{\bm q}=0,0,\ell + 0,0,\delta$,
where $0< \delta < 1$, $\ell$ is an integer, and a factor of $2\pi/{\bar a}$ is implied.
The term $0,0,\delta$ falls in the first zone and, therefore, determines the eigenvalues
and eigenvectors.
Using the values for ${\bm r}^{a}$ and ${\hat z}^{a}$ defined in Table \ref{tab-raza}
we write,
\begin{eqnarray}
{\bm F}_{\perp}^{\alpha}(0,0,\ell+\delta) &=& \frac{(1,1,0)}{\sqrt{3}}
\left( U^{1,\alpha}(\delta)-U^{2,\alpha}(\delta) \right) \\
&+& \frac{(1,-1,0)}{\sqrt{3}}
\left( U^{4,\alpha}(\delta)-U^{3,\alpha}(\delta) \right)\e^{\imath \frac{\ell \pi}{2}}
\nonumber .
\end{eqnarray}
Note that the projections of the spins onto the plane perpendicular 
to the direction of ${\bm Q}$ sum to zero, 
(i.e., ${\hat z}^{(1)}_\perp + {\hat z}^{(2)}_\perp + 
{\hat z}^{(3)}_\perp + {\hat z}^{(4)}_\perp=0$).
For wave vectors ${\bm Q}=0,0,\delta$ and
${\bm Q}=0,0,2+\delta$ one has the following,
\begin{eqnarray}
{\bm F}_{\perp}^{\alpha}(0,0,\delta) &=& \frac{(1,1,0)}{\sqrt{3}}
\left( U^{1,\alpha}(\delta)-U^{2,\alpha}(\delta) \right) \\
&+& \frac{(1,-1,0)}{\sqrt{3}}\left( U^{4,\alpha}(\delta)
-U^{3,\alpha}(\delta) \right) \nonumber ,
\end{eqnarray}
and
\begin{eqnarray}
{\bm F}_{\perp}^{\alpha}(0,0,2+\delta) &=& \frac{(1,1,0)}{\sqrt{3}}
\left( U^{1,\alpha}(\delta)-U^{2,\alpha}(\delta) \right) \\
&-& \frac{(1,-1,0)}{\sqrt{3}}
(U^{4,\alpha}(\delta)-U^{3,\alpha}(\delta)) \nonumber .
\end{eqnarray}
The modulus squared of these two functions, e.g., the numerator of the
scattering cross section, yields the same numerical result,
\begin{eqnarray}
& &|{\bm F}_{\perp}^{\alpha}(0,0,\delta)|^2 =
|{\bm F}_{\perp}^{\alpha}(0,0,2+\delta)|^2 \nonumber \\
\! \! \! &=& \frac{1}{3} \left \{ \left(U^{1,\alpha}(\delta)-U^{2,\alpha}(\delta)-
U^{3,\alpha}(\delta)+U^{4,\alpha}(\delta) \right)^2 \right . \nonumber \\
\! \! \! &+& \left . \left(U^{1,\alpha}(\delta)-U^{2,\alpha}(\delta)+
U^{3,\alpha}(\delta)-U^{4,\alpha}(\delta) \right)^2 \right \} \nonumber.
\end{eqnarray}
This means the scattering cross section, given by 
Eq.~\ref{eq-XsectI}, 
in the limit $\delta \rightarrow 0$, is the same (or exactly 
correlated) for ${\bm Q}=0,0,0$ and ${\bm Q}=0,0,2$, 
absent the magnetic form factor 
$(f(Q))$. Therefore, the paramagnetic scattering of \Tb\ 
can not be generated by a model with Ising spins 
(infinite local \Is\ anisotropy).

In the case of Heisenberg spins with finite single-ion anisotropy, 
we consider the function ${\bm F}_{\mu,\perp}^{\alpha}({\bm q})$ 
is given by Eq.~\ref{eq-FperpH}. Again, restricting ourselves to wave 
vectors along the $(00l)$ direction, 
we have the general result
\begin{eqnarray}
{\bm F}_{\mu,\perp}^{\alpha}(0,0,\ell+\delta)&=&{\bm U}_{\mu,\perp}^{1,\alpha}(\delta) +
{\bm U}_{\mu,\perp}^{2,\alpha}(\delta) \\ 
&+&\left( {\bm U}_{\mu,\perp}^{3,\alpha}(\delta)+{\bm U}_{\mu,\perp}^{4,\alpha}(\delta)
\right) \e^{\imath \frac{\ell \pi}{2}} \nonumber \ ,
\end{eqnarray}
where ${\bm U}_{\mu,\perp}^{a,\alpha}(\delta)=(U_{x,\mu}^{a,\alpha}(\delta),
U_{y,\mu}^{a,\alpha}(\delta),0)$.
For ${\bm Q}$ near $0,0,0$ and $0,0,2$, we obtain the following two forms, 
\begin{eqnarray}
{\bm F}_{\mu,\perp}^{\alpha}(0,0,\delta)&=&{\bm U}_{\mu,\perp}^{1,\alpha}(\delta) +
{\bm U}_{\mu,\perp}^{2,\alpha}(\delta) \\ 
&+& {\bm U}_{\mu,\perp}^{3,\alpha}(\delta)+{\bm U}_{\mu,\perp}^{4,\alpha}(\delta)
\nonumber \ ,
\end{eqnarray}
and
\begin{eqnarray}
{\bm F}_{\mu,\perp}^{\alpha}(0,0,2+\delta)&=&{\bm U}_{\mu,\perp}^{1,\alpha}(\delta) +
{\bm U}_{\mu,\perp}^{2,\alpha}(\delta) \\
&-& \left( {\bm U}_{\mu,\perp}^{3,\alpha}(\delta)+{\bm U}_{\mu,\perp}^{4,\alpha}(\delta) \right)
\nonumber  \ .
\end{eqnarray}
Taking the modulus squared we get,  
\begin{equation}
\label{eq-FH0}
|{\bm F}_{\mu,\perp}^{\alpha}(0,0,\delta)|^2=(A^2+B^2)+(C^2+D^2)+2(AC+BD)\ ,
\end{equation}
and
\begin{equation}
\label{eq-FH2}
|{\bm F}_{\perp}^{\alpha,\mu}(0,0,2+\delta)|^2=(A^2+B^2)+(C^2+D^2)-2(AC+BD)\ ,
\end{equation}
where
\begin{eqnarray}
A&=&U_{x,\mu}^{1,\alpha}(\delta)+U_{x,\mu}^{2,\alpha}(\delta) \ , 
\nonumber \\
B&=&U_{y,\mu}^{1,\alpha}(\delta)+U_{y,\mu}^{2,\alpha}(\delta) \ ,
 \\
C&=&U_{x,\mu}^{3,\alpha}(\delta)+U_{x,\mu}^{4,\alpha}(\delta) \ ,
\nonumber \\
D&=&U_{y,\mu}^{3,\alpha}(\delta)+U_{y,\mu}^{4,\alpha}(\delta) \ .
\nonumber
\end{eqnarray}
Equations \ref{eq-FH0} and \ref{eq-FH2} are not strictly equivalent. 
Hence, it is possible to have paramagnetic spin-spin correlations 
about ${\bm Q}=0,0,2$ while intensity about ${\bm Q}=0,0,0$ 
is suppressed. This result 
puts on a firm theoretical footing the need to describe \Tb\ by a 
three-component Heisenberg model with finite anisotropy.  

\vfill

%\newpage %Just because of unusual number of tables stacked at end

\newpage

\newpage

\bibliography{pmscatt-expt,pmscatt-thry,pmscatt-comments}% Produces the bibliography via BibTeX.

\end{document}